\documentclass[10pt,a4paper]{article}

\usepackage[utf8]{inputenc}
\usepackage{amsmath, amsthm}
\usepackage{amsfonts, dsfont}
\usepackage{amssymb}
\usepackage{graphicx}
\usepackage[ruled,vlined]{algorithm2e}
\usepackage{caption,subcaption}
\usepackage{color}
\usepackage{comment}
\usepackage{mathtools}
\mathtoolsset{showonlyrefs}

\newcommand{\R}{\mathbb{R}}
\newcommand{\e}{\varepsilon}
\newcommand{\ep}{\varepsilon}

\newcommand{\dt}{{\Delta t}}
\newcommand{\dx}{{\Delta x}}
\newcommand{\dz}{{\Delta z}}

\newtheorem{theorem}{Theorem}
\newtheorem{statement}[theorem]{Statement}

\definecolor{ln}{RGB}{253,106,2}

\definecolor{amethyst}{rgb}{0.6, 0.4, 0.8}

\definecolor{Sam}{RGB}{1, 58, 223}

\definecolor{Anya}{RGB}{0,170,0}

\usepackage[comma]{natbib}
\begin{document}
\author{Vincent Calvez\footnote{CNRS \& Institut Camille Jordan, Lyon.  \texttt{Vincent.Calvez@math.cnrs.fr}}, 
Susely Figueroa Iglesias\footnote{Institut de Math\'ematiques de Toulouse; UMR 5219, Universit\'e de Toulouse, UPS, IMT, F-31062 Toulouse Cedex 9, France. e-mail: \texttt{Susely.Figueroa@math.univ-toulouse.fr}}, 
Hélène Hivert\footnote{Univ Lyon, École centrale de Lyon, CNRS UMR 5208, Institut Camille Jordan, F-69134 Écully, France. e-mail: \texttt{helene.hivert@ec-lyon.fr}},  \\
Sylvie M\'el\'eard\footnote{CMAP, UMR 7641, Ecole polytechnique, France. \texttt{sylvie.meleard@polytechnique.edu}}, Anna Melnykova\footnote{Universit\'e de Cergy-Pontoise, AGM CNRS UMR 8088 and Universit\'e de Grenoble Alpes, LJK CNRS UMR 5224, e-mail: \texttt{anna.melnykova@u-cergy.fr}}, Samuel Nordmann\footnote{CAMS, EHESS, PSL Université, Paris, e-mail: \texttt{samuel.nordmann@ehess.fr}}}
\date{}

\title{
Horizontal gene transfer: numerical comparison between stochastic and deterministic approaches}

\maketitle

\begin{abstract} Horizontal gene Transfer (HT) denotes the transmission of genetic material between two living organisms, while the vertical transmission refers to a DNA transfer from parents to their offspring. Consistent experimental evidence report that this phenomenon plays an essential role in the evolution of certain bacterias. In particular, HT is believed to be the main instrument of developing the antibiotic resistance. In this work, we consider several models which describe this phenomenon: a stochastic jump process (individual-based) and the deterministic nonlinear integrodifferential equation obtained as a limit for large populations. We also consider a Hamilton-Jacobi equation, obtained as a limit of the deterministic model under the assumption of small mutations. The goal of this paper is to compare these models with the help of numerical simulations. More specifically, our goal is to understand to which extent the Hamilton-Jacobi model reproduces the qualitative behavior of the stochastic model and the phenomenon of evolutionary rescue in particular.\end{abstract}


\section*{Introduction}

Accurate mathematical description of the evolutionary mechanism is an open question in biology, medicine, and industry. In particular, transmission of pathogens, or antibiotic resistance of bacteria is directly linked to the ability of the bacteria population to mutate and exchange genetic material either vertically (from parents to offspring), or horizontally (from the interaction between non-parental individuals). 

Horizontal Gene Transfer was first described in bacteria when the antibiotic resistance was discovered. This resistance occurs when one bacterial cell becomes resistant to an antibiotic due to mutation, and then transfers resistance genes to other species of bacteria. 
However the Horizontal Transfer of biologic information is not restricted to genes, it also describes the transfer of plasmids and endosymbionts, see for example \cite{Henry-etal, Lili399}. 
Some artificial applications of horizontal transfer include forms of genetic engineering (Gene Delivery) that result in an organism with its genes changed in some way, and, consequently, possessing new properties or functions (see for instance \cite{GeneDelivery}). These applications are particularly useful for "Gene Therapy", which is an experimental procedure that may help treat or prevent genetic disorders and some types of cancer.

The primary goal of our work is to describe the mechanism of the transfer itself and explain how it affects the population dynamics.  {Throughout the paper we abbreviate the Horizontal Transfer to HT.}

Our study starts with finding a good model of a bacteria population. Several mathematical models for describing a population dynamics were proposed in literature. The first model we consider is a stochastic birth and death process (see, for reference, \cite{BFRMT_Stochastic, Fournier&Meleard}), which describes the dynamics of reproduction, competition, and exchange of genetic material between individuals in a population. The phenotype of each individual is described by a numerical parameter, called trait. Numerical experiments show that the effect of a unilateral horizontal gene transfer may lead to a cyclic behavior of the population. Roughly speaking, while HT drives individuals towards a non-fit phenotype --- and, consequently, to extinction, very few not affected by transfer fit individuals may eventually repopulate the environment, before being driven again to deleterious phenotypes. This phenomenon is called an \emph{evolutionary rescue of a small population}. 

However, within a framework of stochastic jump processes, it is hard to define and study the observed cycling phenomena accurately. The second drawback of the stochastic system is that it is costly to compute, especially for a large time scale and population size. Thus, in the case of a large population, it is more practical to work with a deterministic PDE model, describing  {the limiting behaviour of a stochastic system when the population size goes to infinity} \cite{Billiard2016, BFRMT_Effect, FRT2009}.
In certain settings, the population dynamics involve concentration phenomena (i.e., the convergence of the population density to singular solutions, such as Dirac masses). In that case, the PDE formulation is not suitable. Thus, applying a limiting procedure for small mutations and time rescaling to the PDE model, we pass to a Hamilton-Jacobi type equation. 

The primary goal of our work is thus to conduct a numerical analysis of the population dynamics on a macroscopic individual-based model and to compare it with the deterministic system which is obtained as a limit for a large population. We are especially interested in determining to which extent the limiting Hamilton-Jacobi equation can grasp qualitative properties of the stochastic model.
This framework has already been successfully used to understand the concentration phenomena, and the location of the dominant trait (see for instance \cite{LMP2011, Perthame_Barles2008}). We aim to understand if the Hamilton-Jacobi approach is also well suited to describe the evolutionary rescue phenomena which crucially rely on an accurate description of the small populations.

On this step, the choice of an approximation scheme for simulating solutions of the PDE model is of tremendous importance. As we further explain in Section \ref{section:numerical_tests}, classical explicit schemes do not preserve the asymptotic behavior of the solution if the time rescaling step goes to $0$. From a numerical point of view, it involves operations with exponentially big values, which lead to non-negligible errors for explicit numerical schemes. We address this question by proposing an asymptotic preserving scheme for a Hamilton-Jacobi equation, adapting an approach proposed in \cite{Grandall1984}. More generally, the numerical approximation problem for solutions of Hamilton-Jacobi equations is treated in \cite{Barles_Ishii}.

This paper is structured as follows: in Section \ref{section:model} we introduce the model both in a stochastic and deterministic setting, and formally derive the limiting Hamilton-Jacobi equation. Then, we simulate a jump process, describing the bacteria population, and study its properties for different values of parameters. Numerical experiments are gathered in Section \ref{section:numerical_tests}. We aim to numerically determine the critical HT rate, which leads to an almost sure extinction of the whole population. On the next step, we conduct the same analysis for a Hamilton-Jacobi equation with the help of an asymptotic preserving scheme and compare it with the stochastic model on an appropriate timescale, and explain why the classical scheme fails to work. We end our study with conclusions and discussion of yet unsolved numerical and theoretical questions.

\section{Model}\label{section:model}
\subsection{Stochastic model}
We consider a stochastic model describing the evolution of a population structured by phenotype. In a general case it is described at each time $t$ by the point measure
\begin{equation}\label{eq:point_measure}
\nu^K_t(dx) = \frac{1}{K}\sum_{i=0}^{N^K_t}\delta_{X_i(t)}(dx),
\end{equation}
where parameter $K$ is a scaling parameter, referred to as the \emph{carrying capacity}. It stands for the maximal number of individuals that the underlying environment is able to host ($K$ can represent, for example, the amount of available resources). $N^K_t = K\int \nu^K_t(dx)$ is the size of the population at time $t$,  {and $X_i(t)\in\mathds{R}^n$ is the trait of $i$-th individual living at $t$, which summarizes the phenotype information. In this work we assume $n=1$, that is, the trait is given by a point on a real line. }

The demography of the population is regulated, first of all, by its birth and death rates. An individual with a trait $x$ gives birth to a new individual with rate $b(x)$.  {The trait $y$ of the offspring is chosen from a probability distribution $m(x-y)dy$ (by that we mean that  $\int_{\mathds{R}} m(x-y)dy = 1$). We will refer to it as the \emph{mutation kernel}. }
An individual with a trait $x$ dies according to an intrinsic death rate $d(x)$ plus an additional death rate $C\dfrac{N_t^K}{K}$ (independent of $x$) which stands for the competition between individuals. 

Finally, an individual with a trait $x$ can induce a \emph{unilateral} HT to an individual with trait $y$ at rate $h_K(x,y,\nu)$, such that the pair $(x,y)$ becomes $(x,x)$.  {In literature this kind of transfer is sometimes referred to as a \emph{conjugation}. } For simplicity, we assume $h_K(x,y,\nu)$ to be in the particular form 
\begin{equation}\label{FormOfh}
h_K(x,y,\nu) = h_K(x-y,N) =\tau_0\frac{\alpha(x-y)}{N/K},
\end{equation}
where $N=K\int_{\mathds{R}}\nu(dx)$ is the number of individuals, $\tau_0>0$ is a constant and $\alpha$ is either a  Heaviside, or a smooth bounded function, such that for a small $\delta>0$:
\begin{gather}\label{eq:alpha}
\quad \alpha(z)=
\left\{\begin{aligned}
    &0 &&\text{if }z<-\delta\\
    &1 &&\text{if }z>+\delta
    \end{aligned}
\right.,
\quad \alpha'(0)=\frac{1}{2\delta},
\end{gather}
where $\delta$ is the stiffness parameter. We introduce $\delta$ to have the advantage of working with a smooth function (which will be useful in the following parts), while mimicking the binary nature of the Heaviside function. 

For a population $\nu=\frac{1}{K}\sum_{i=1}^N\delta_{x_i}$ and a generic measurable bounded function $F$, the generator of the process is then given by: 
\begin{multline*}
L^K F(\nu) = \sum_{i=1}^N b(x_i)\int_{\mathds{R}} \left(F\left(\nu+\frac{1}{K}\delta_{y}\right) - F(\nu)\right)m(x_i, dy)\\
+\sum_{i=1}^N \left(d(x_i) + C\frac{N}{K} \right)\left(F\left(\nu-\frac{1}{K}\delta_{x_i}\right) - F(\nu)\right) \\ 
+\sum_{i,j=1}^N h_K(x_i, x_j, \nu)\left( F\left(\nu + \frac{1}{K}\delta_{x_i}-\frac{1}{K}\delta_{x_j} \right) - F(\nu) \right).
\end{multline*}

It is standard to construct the measure-valued process $\nu^K$ as the solution of a stochastic differential equation driven by Poisson point measures and to derive moment and martingale properties (see for instance \cite{Fournier&Meleard}).

\subsection{The PDE model}\label{subsection:pde_model}
It is proven (see in particular \cite{Billiard2016, CFM2008}) that for $K\to+\infty$ the stochastic process defined by a sequence of point measures given by \eqref{eq:point_measure} converges  {in probability to the unique solution of a non-linear integro-differential equation.} This equation is given by:
\begin{equation*}\label{equation:PDE}
\left\{
\begin{array}{rcl}
\partial_t f(t,x) & = & -(d(x)+C\rho_1(t))f(t,x)+ \displaystyle\int_{\R^n} m(x-y)b(y)f(t,y)dy+\\ 
& & \qquad f(t,x)\int_{\R^n}\tau(x-y)\frac{f(t,y)}{\rho_1(t)}dy, \quad (t,x) \in \R_+\times\R^n,\\
\rho_1(t)& = &\displaystyle\int_{\mathds{R}} f(t,x)dx,\\
f(0,x)& = &f^0(x)>0,
\end{array}
\right.
\end{equation*}
where $f(t,x)$ is the macroscopic density of the population with trait $x$ at time $t$ and, accordingly to the previous section, $b(x)$, $d(x)$ and $C$ are the birth, death and competition rate respectively, $m$ is the mutation kernel, and 
\begin{equation}
\label{eq:tau_def}
\tau(y-x):=\tau_0\left[\alpha(x-y)-\alpha(y-x)\right]
\end{equation} 
is the horizontal transfer flux.

Now our goal is to pass from micro- to a macroscopic scale with the help of a rescaling. 
On the one hand, we consider the case of small mutations: for a small parameter $\e>0$ we define
$$m_\e(x-y)=\frac{1}{\e^n}m\left(\frac{x-y}{\e}\right).$$
With a change of variable $z=\frac{x-y}{\e}$ we can rewrite the mutation term at $(t,x)$ as
\begin{equation*}\label{mut_renorm}
\int_{\R^n} m_\e(x-y)b(y)f(t,y)dy= \int_{\R^n} m(z)b(x+\e z)f(t,x+\e z)dz.
\end{equation*}
On the other hand, when $\e$ is small, the effect of mutations can only be observed in a larger time scale. Thus, we rescale time with $t\mapsto\frac{t}{\e}$.

We end up with the following system, for $\e>0$, and $(t,x)\in \R_+\times\R^n$:
\begin{equation}\label{equation:PDE_eps}
\left\{
\begin{array}{rcl}
\e \partial_t f_\e(t,x) &=& -(d(x)+C\rho_\e(t))f_\e(t,x)+ \int_{\R^n} m(z)b(x+\e z)f_\e(t,x+\e z)dz+\\ 
& & \qquad f_\e(t,x)\int_{\R^n}\tau(x-y)\frac{f_\e(t,y)}{\rho_\e(t)}dy,\\
\rho_\e(t)& = &\displaystyle\int_{\mathds{R}} f_\e(t,x)dx,\\
f_\e(0,x)& = & f_\e^0(x)>0.
\end{array}
\right.
\end{equation}

\subsection{The Hamilton-Jacobi limit}\label{subsection:hamilton-jacobi}
We now derive the limiting problem \eqref{equation:PDE_eps} when $\e\to0$. As we will see, the limiting problem allows us to give a rigorous mathematical framework and to perform useful formal calculations.

Equations in the form of \eqref{equation:PDE_eps} often give rise to a concentration phenomenon, i.e the convergence of $f_\e$ towards a Dirac mass when $\e\to0$ (see \cite{Perthame_Barles2008,DJMP}). The usual way to deal with these asymptotics is to perform a Hopf-Cole transformation (or WKB ansatz), i.e to consider 
\begin{equation}\label{eq:u_e_def}
u_\e(t,x):= \e \ln(f_\e(t,x)).
\end{equation}
This change of variable comes from the intuition that a Dirac mass is no more than a narrow Gaussian, and more precisely that $f_\e$ should behave like a Gaussian of variance $\e$ when $\e\to0$.
Accordingly, we expect $u_\e$ to have a non singular limit when $\e\to0$. Incidentally, this substitution also gives insights on the convenient scheme to use for numerical simulations,  as we will see in the following section.

Now our goal is to identify and derive the asymptotic properties of $u_\e$ when $\e\to0$, which will be used for discussions in the sequel. The following computations are only formal, since rigorous proofs are often intricate in this context.
Substituting \eqref{eq:u_e_def} into \eqref{equation:PDE_eps} we deduce that $u_\e$ satisfies 
\begin{multline}\label{equation:hamilton_jacobi}
\partial_t u_\e = -(d(x)+C\rho_\e(t))+\int_{\R^n} m(z)b(x+
\e z)\exp\left\{\frac{u_\e(t,x+\e z)-u_\e(t,x)}{\e}\right\}dz\\ +\int_{\R^n}\tau(x-y)\frac{f_\e(t,y)}{\rho_\e(t)}dy.
\end{multline}

Formally, at the limit $\e\to0$, $u_\e$ converges to a continuous function $u$ which satisfies the following Hamilton-Jacobi equation in the \textit{"viscosity"} sense:
\begin{equation}
\label{HJ_limit}
\partial_t u = -(d(x)+C\rho(t))+b(x)\int_{\R^n} m(z)e^{z\cdot\nabla_x u}dz+\tau(x-\overline{x}(t)),
\end{equation}
where 
$\rho(t)\geq0$ is the weak limit of $\rho_\e(t)$ and 
\begin{equation}\label{definition_x_bar}
\bar x(t)=\mathrm{argmax}\ u(t,\cdot).
\end{equation} 
We formally assume here and in the following that the definition of $\bar x(t)$ is unambiguous, i.e that $u$ reaches its maximum in a single point.
Note that the limiting function $u$ is not expected to be $C^1$ for all time. We thus need to deal with a generalized notion of solutions, namely \emph{viscosity solution} (see \cite{GuyBarles}).

This framework is convenient because most of the information is contained in the dynamics of $\bar x(t)$. See the next section for further analysis.

\subsection{Formal analysis on the Hamilton-Jacobi equation}\label{sec:adaptive_dynamics}
Hamilton-Jacobi equations are particularly known in mathematical biology to be a good model to describe how a population concentrates around the dominant trait(s) when the mutations are small. However, here we are interested to use this model to describe a phenomenon of \emph{evolutionary rescue}. In this subsection we attempt an analysis of the equation \eqref{HJ_limit}. We point out that the calculations are only formal, since rigorous proofs are intricate and beyond the scope of this paper.

\subsubsection{Generality}
From an integration of \eqref{equation:PDE_eps} with respect to $x$ and classical computations (under the assumptions of bounded functions for the birth, death and transfer rates), we deduce that our model satisfies a \emph{saturation property}, i.e. $\rho_\e(t)$ is bounded from above, uniformly in $t\geq0$ and $\e>0$.
From this and $\rho_\e(t)=\int_{\R^n}e^{\frac{u_\e(t,x)}{\e}}dx$, we deduce that
$\forall t>0$, $\sup\limits_{x\in\R^n} u(t,x)\leq 0$ and the following constraint holds:
\begin{equation}
\label{constraint_HJ}
\sup\limits_{x\in\R^n} u(t,x)=0\quad\text{when}\quad \rho(t)>0.
\end{equation}
Note that our model allows the population to get extinct, thus we cannot expect $\rho$ to be always positive. 
As a byproduct, we derive the \textit{concentration property}, i.e the formal weak convergence of measures
\begin{equation*}\label{formal_conv}
f_\e(t,x)\rightharpoonup\rho(t)\delta_{\bar x(t)}(dx),\quad\text{when }\e\to 0,
\end{equation*}
where $\delta_{\bar x(t)}$  denotes, as usually, the Dirac measure centered in $\bar x(t)$.
From \eqref{constraint_HJ} it is possible to formally derive a formula for $\rho$. Indeed, either $\rho(t)=0$ or $\rho(t)>0$ and
\begin{equation*}\label{dt_u_xvar}
\partial_t u(t,\bar x(t))=0,
\end{equation*}
which implies
\begin{equation}\label{formula_rho}
\rho(t)=\frac{b(\bar x(t))-d(\bar x(t))+\tau(0)}{C}=\frac{b(\bar x(t))-d(\bar x(t))}{C},
\end{equation}
for $\tau$ defined in \eqref{eq:tau_def}.

Having above definitions in hand, we can now perform a formal analysis on the dynamics of $\bar x(t)$, defined below in \eqref{Assumption_x_bar}. Our aim is to show how the behaviour of the system can be analyzed within the framework of a Hamilton-Jacobi equation \eqref{HJ_limit}. To fix ideas, we fix all constants but $\tau_0$ and we assume \eqref{eq:birth_rate}-\eqref{eq:mutation-kernel} as follows:
\begin{gather}
b(x)=b_r>0, \label{eq:birth_rate} \\
d(x)=d_r x^2,\quad d_r>0, \label{eq:death_rate}\\
m(z)=\frac{1}{\sqrt{2\pi}\sigma}e^{-\frac{z^2}{2\sigma^2}}, \label{eq:mutation-kernel}
\end{gather}
and the transfer function $h_K(x,y,\nu)$ is defined in \eqref{FormOfh}.
Moreover we work under the following assumptions:
\begin{equation}\label{Assumption_x_bar}
\begin{aligned}
& u(t,\cdot) \text{ reaches its maximum on a single point }\bar x(t),\\
& \bar x(t) \text{ is a non-degenerate maximum, i.e }\nabla_x^2u(t,x)<0,\\
& \bar x(t) \text{ is smooth with respect to }t.
\end{aligned}
\end{equation}
Finally we assume that the initial condition $f^0$ is a given function of $x$ which reads:
\begin{equation}\label{eq:init_condition_f}
f_\e^0(x) = \frac{1}{\sqrt{\e}}e^{-\frac{x^2}{2\e}}.
\end{equation}

\subsubsection{Smooth dynamics $\bar x(t)$.}
The following statement deals with the smooth dynamics of $\bar x(t)$, i.e in the regime where no jump occurs in the dynamics of $\bar x(t)$. 
\begin{statement}\label{Dynamics_x_bar}
Under assumptions \eqref{eq:birth_rate}-\eqref{Assumption_x_bar}, the function $t\mapsto \bar{x}(t)$ is an increasing function which satisfies the following inequality $\forall t\geq 0$:
$$
0\leq \bar x(t)\leq \frac{\tau_0}{2d\delta}.
$$
More precisely, $\bar x(t)$ satisfies the {canonical equation} \begin{equation}\label{CanonicalEquation}
\frac{\mathrm{d}}{\mathrm{d}t}\bar x(t)= \left[-\nabla^2_x u(t,\bar x(t))\right]^{-1}\cdot\left(\nabla_x r(\bar x(t))+\nabla_x\tau(0)\right),
 \end{equation}
where
\begin{equation}\label{Definition_r}
r(x):=b(x)-d(x),
\end{equation}
and $\nabla^2_x u$ denotes the Hessian of $u$ with respect to the $x$ variable.
\end{statement}
\begin{proof} Under the above assumptions we can derive the dynamics of $\bar x(t)$, referred to as the \emph{canonical equation} in the literature (see for instance \cite{Mirrahimi-Roquejoffre-Cras}). Indeed, starting from
\begin{equation*}\label{grad_u_xvar}
\nabla_x u (t,\bar x(t))=0,
\end{equation*}
a differentiation with respect to $t$ gives~\eqref{CanonicalEquation}.
Equation \eqref{CanonicalEquation} has a unique singular point $x_\star$, which satisfies $r'(x_\star)+\tau'(0)=0$, with $\tau$ defined in \eqref{FormOfh} and $r$ in \eqref{Definition_r}. 
We find
\begin{equation}\label{Definition_x_star}
x_\star= \frac{\tau_0}{2d_r\delta}.
\end{equation} Note that $t\mapsto \bar x(t)$ is increasing when $\bar x(t)<x_\star$ and decreasing when $\bar x(t)>x_\star$. Besides, from the initial condition~\eqref{eq:init_condition_f}, we have $\bar x(0)=0$,
and consequently $0\leq \bar x(t)\leq x_\star \quad \forall t$. \end{proof}

\subsubsection{Evolutionary rescue.}

In general, the canonical equation~\eqref{CanonicalEquation} does not hold in every point of time. Indeed, a new maximum of $u$ can arise in a finite time, which would cause a "jump" in the dynamics of $\bar x(t)$: this is what we call an \emph{evolutionary rescue}. Formally, this is what happens (periodically in time) in the case of cycles, see Figure~\ref{fig:cycles_HJ}. We thus expect $\bar x(t)$ to possibly jump periodically, and to follow \eqref{CanonicalEquation} between two jumps. We now try to characterize the possible jumps.
For $T>0$, we denote 
$$\bar x(T^-):=\lim\limits_{\substack{t\to T\\ t<T}}\bar x(t),\quad \bar x(T^+):=\lim\limits_{\substack{t\to T\\ t>T}}\bar x(t).$$

\begin{statement}\label{Jumps_Characterisation}
We assume that \eqref{eq:birth_rate}-\eqref{Assumption_x_bar} hold until a time $T>0$, such that $u(T,\cdot)$ reaches its maximum on $\bar x(T^-)$ and on another point $\tilde x$. Then $\tilde x=0$ and $\bar x(t)$ will jump towards $0$ at time $T$, i.e $\bar x(T^+)=0$. 
\end{statement}
\begin{proof}
From assumption~\eqref{Assumption_x_bar}, we have  $\forall t\in[0,T]$ that $u(t,\cdot)$ is concave non-degenerate on $[\bar x(t)\pm \theta],$ with $\theta>0$. For simplicity, we further assume $\delta\leq\theta$, where $\delta$ is defined in~\eqref{eq:alpha}. 

First, let us show that $\tilde x=0$. We define the fitness function of trait $x$ in a population concentrated in $\bar x$:
\begin{equation*}
F_{\bar x}(x):= r(x)+\tau(x-\bar x),
\end{equation*}
where $r$ and $\tau$ are respectively defined in \eqref{Definition_r} and \eqref{eq:tau_def}.
Note that we have $\partial_t u(t,x)=F_{\bar x(t)}(x)-C\rho(t)$, for $t<T$. 
But $\tilde x\not\in [\bar x(t)\pm \delta]$ and the choice of parameters \eqref{eq:birth_rate}-\eqref{eq:death_rate}-\eqref{eq:alpha} implies $\tilde x$ must maximize $F_{\bar x(T^-)}(\cdot)$, hence $\tilde x=0$.

The second step is to prove that there will be an actual jump towards $0$, i.e $\bar x(T^+)=0$. First, note that there exists a small $\eta>0$ such that $\forall t\in(T-\eta,T)$, $u(t,\bar x(t))=0$ and $u(t,0)<0$. Let us fix $t\in(T-\eta,T)$. 
We have $F_{\bar x(t)}(0)\geq F_{\bar x(t)}(\bar x(t))$, and we claim that the inequality is strict. Indeed, since $t\mapsto x(t)$ is increasing, $F_{\bar x(t)}(\bar x(t))$ is decreasing, whereas $F_{\bar x(t)}(0)$ is constant (as long as $\eta$ is small enough such that $\bar x(T-\eta)>\delta$). We end up with 
\begin{equation*}
F_{\bar x(t)}(0)> F_{\bar x(t)}(\bar x(t)).
\end{equation*}
The above inequality expresses the fact that $0$ is fitter than $\bar x(t)$ in a population with trait $\bar x(t)$. In general, this does not allow to conclude that $0$ will invade and become the new dominant trait (i.e., that the jump will occur) because it does not imply that $0$ will remain fitter during all the process of invasion.
But the particular form of our problem, especially the fact that $\tau$ is an odd function, implies 
\begin{equation*}
F_{0}(0)> F_{0}(\bar x(t)).
\end{equation*}
Indeed we have from the definition of $F_{\bar x}(x)$ that
$$
F_{0}(0)- F_{0}(\bar x(t))=r(0)-r(\bar x(t))+\tau(\bar x-\bar x)-\tau(0)=d_r \bar x(t)^2>0.
$$
Consequently that for all $\lambda\in[0,1]$
$$\lambda F_{0}(0)+(1-\lambda)F_{\bar x(t)}(0)>\lambda F_{0}(\bar x(t))+(1-\lambda)F_{\bar x(t)}(\bar x(t)).$$
It shows that $0$ remains the fittest trait during all the process of invasion, and therefore that $0$ will actually invade, i.e that $\bar x(t)$ will actually jump towards $0$ at time $T^+$. 
\end{proof}

\subsubsection{Threshold for cycles}
In the previous section, we described the possible \emph{evolutionary rescue}, i.e the possible jumps in the dynamics of $\bar x(t)$ towards $x=0$. 
When a jump occurs, a new cycle begins: it leads to a periodical behavior of $\bar x(t)$, hence the cycling phenomenon.

We recall that a jump corresponds to a rescue of the population concentrated at $\bar x(t)$ by the small population with trait $x=0$. It is possible only if $\bar x(t)>\delta$ and if $0$ is fitter than $\bar x(t)$ during a sufficiently large interval of time (which is the time needed for the small population at $x=0$ to regrow). Note that $0$ is fitter than $\bar x(t)$ if and only if
\begin{align}\label{Definition_x_resc}
F_{\bar x(t)}(0)\geq F_{\bar x(t)}(\bar x(t))\quad&\text{iff}\quad b_r-\tau_0\geq b_r-d_r\bar x(t)^2,\\
&\text{iff} \quad\bar x(t)\geq x_{resc}:=\sqrt{\frac{\tau_0}{d_r}}.
\end{align}
But if no jump occurs, $\bar x(t)$ formally follows \eqref{CanonicalEquation}, thus $\bar x(t)< x_\star$ and $\bar x(t)$ converges to $x_\star$ when $t\to+\infty$ (with $x_\star$ is defined in \eqref{Definition_x_star}). 
\begin{statement}
Under assumptions \eqref{eq:birth_rate}-\eqref{Assumption_x_bar},
 the evolutionary rescue phenomena occurs if and only if 
\begin{equation} \label{definition_tau_cyc}
 \tau_0>\tau_{cyc}:= 4d_r\delta^2.
 \end{equation}
\end{statement}
 {Note that the condition $\tau_0>\tau_{cyc}$ is equivalent to $x_{resc} < x_\star$ , which are defined respectively in \eqref{Definition_x_star} and \eqref{Definition_x_resc}.}

\subsubsection{Threshold for extinction.}
The population is said to be "extinct" at time $t$ if $\rho(t)=0$. 
According to \eqref{formula_rho}, we define $x_{ext}$ as to solve $r(x_{ext})=0$, i.e
\begin{equation}\label{Definition_x_ext}
x_{ext}:= \sqrt\frac{b_r}{d_r},
\end{equation}
that is, a population concentrated at trait $\bar x$ is extinct iff $\bar x\geq x_{ext}$.

The picture is simple in the case of stabilization without cycles, i.e when $\tau_0\leq\tau_{cyc}$ (see \eqref{definition_tau_cyc}). In this case, we recall that $\bar x(t)$ formally follows \eqref{CanonicalEquation} for all $t>0$, thus $\bar x(t)< x_\star$ and $\bar x(t)$ converges to $x_\star$ when $t\to+\infty$ (where $x_\star$ is defined in \eqref{Definition_x_star}). Thus, if $x_\star\leq x_{ext}$, we have $\rho(t)>0$ for all $t>0$; on the contrary, if $x_\star> x_{ext}$, there exists a time $t_{ext}>0$ for which $\rho(t)=0$ for all $t\geq t_{ext}$. It gives a sharp threshold for extinction of the population: indeed, the population eventually gets extinct if and only if $x_\star>x_{ext}$, which naturally leads us to the following statement. 
\begin{statement} Under assumptions \eqref{eq:birth_rate}-\eqref{Assumption_x_bar}, if $\tau_0\leq\tau_{cyc}$, then 
 the population eventually gets extinct if and only if 
 \begin{equation}\label{eq:birth_death_formal}
\tau_0> \tau_{ext}:=2\sqrt{b_rd_r}\delta.
\end{equation}
\end{statement}

We point out that, surprisingly enough, $\tau_{ext}$ is an increasing function of the death rate $d_r$, meaning that under a higher death rate, the population can survive to a higher HT rate. The interpretation we propose is that if $d_r$ is high, the population driven outward $x=0$ dies rapidly, thus the  population that remained closer to $0$ undergoes a milder HT, which makes the overall population more resistant to a high HT rate. 

Let us now focus on the case where the cycling phenomenon occurs, i.e when $\tau_0>\tau_{cyc}$. In this case, $\bar x(t)$ will follow \eqref{CanonicalEquation} and will periodically jump to $x=0$. First, note that if $x_\star<x_{ext}$, $\bar x(t)$ remains below $x_{ext}$ for all $t$ and the population does not get extinct:
\begin{equation*}
\text{if}\quad\tau\leq \tau_{ext},\quad\text{then}\quad \rho(t)>0,\quad \forall t>0.
\end{equation*}
The most intricate case is when $x_\star>x_{ext}$, which contains cases of extinction and non-extinction, depending on whether the jump of $\bar x(t)$ towards $0$ happens before or after $\bar x(t)$ has passed beyond $x_{ext}$. In other words, extinction can be avoided if the evolutionary rescue happens before the dominant trait is led to extinction, i.e  if $\bar x(T^-)\leq x_{ext}$, where $T$ is the time where the jump of $\bar x(t)$ towards $0$ occurs. However, we are not able to give a satisfactory formula or estimate on $T$. 

Besides, when the jump of $\bar x(t)$ occurs, it can happen that the trait $x=0$ is not fit enough to avoid extinction:  in this case the evolutionary rescue does not manage to sustain the population. It corresponds to the case $x_{resc}>x_{ext}$. We have the following threshold:
the evolutionary rescue is able to sustain the population iff $r(0)+\tau_0>0$, which is equivalent to
\begin{equation}\label{Definition_tau_sus}
\tau_0<\tau_{sus}:=b_r.
\end{equation}

If $\tau\geq \tau_{sus}$, the population eventually gets extinct. If $\tau<\tau_{sus}$, the population is effectively rescued by the evolutionary rescue, even in the case where it passed through an episode of extinction during the previous cycle: in some cases the population is able to regrow after being extinct, which can be seen on Figure~\ref{fig:extinction_HJ}. We think this is an interesting feature that the Hamilton-Jacobi approach is able to grasp. Regarding the stochastic model, an episode of extinction on Hamilton-Jacobi corresponds to an interval of time where the population reaches extremely small values (of order $e^{-\frac{1}{\e}}$, with $\e$ the variance of the mutation kernel), and the probability that every individual dies is bigger than the survival of the population.  

\begin{statement}Assume \eqref{eq:birth_rate}-\eqref{Assumption_x_bar} and $\tau_0>\tau_{cyc}$.
\begin{itemize}
\item if $\tau_0\leq \tau_{ext}$, the population never gets extinct.
\item the evolutionary rescue effectively manages to sustain the population if and only if $\tau_0 <\tau_{sus}:=b_r$. 
\end{itemize}
\end{statement}

\subsubsection{Characteristics of a Hamilton-Jacobi equation}\label{sec:characteristics}
Denoting
\begin{equation*}
-H(t,x,p):= -(d(x)+C\rho(t))+b(x)\int_{\R}m(z)e^{pz}dz+\tau(x-\bar x(t)),
\end{equation*}
from \eqref{HJ_limit} we have $\partial_t u(t,x)+H(t,x,\nabla_x u(t,x))=0$. Since $H$ is convex in the $p$ variable, we have the following representation formula (see \cite{Lions}).
\begin{equation}\label{RepresentationFormula}
u(t,x)=\inf\limits_{\substack{\gamma\in C^0(\R_+,\R)\\ \gamma(t)=x}}\left[\int_0^t L\left(s,\gamma(s),\dot\gamma(s)\right)ds+u^0(\gamma(0))\right],
\end{equation}
where $L(t,x,v)$ is the Lagrangian of the equation, obtained through a Legendre transform (or a convex conjugate) of~$H$.

Every $\gamma$ which is admissible as a minimizer in~\eqref{RepresentationFormula} is called a \emph{characteristic} of the Hamilton-Jacobi equation~\eqref{HJ_limit}. 
Note that every characteristic $\gamma$ formally satisfies the condition
\begin{equation}\label{Equation_Characteristics}
\frac{\mathrm{d}}{\mathrm{d}s}\left[\partial_v L\left(s,\gamma,\dot\gamma(s)\right)\right]=\partial_x L\left(s,\gamma(s),\dot\gamma(s)\right).
\end{equation}
 {\eqref{Equation_Characteristics} holds because $\gamma$ is a critical point of the functional defined in~\eqref{RepresentationFormula}. 
Note that if we replace $H$ by $\tilde H(x,p)= -\frac{x^2}{2}+\frac{p^2}{2}+1,$
the Legendre transform of $\tilde H$ can be computed explicitly:}
\begin{equation*}
\tilde L(x,v)=\frac{x^2}{2}+\frac{v^2}{2}-1.
\end{equation*}
Then~\eqref{Equation_Characteristics} becomes
\begin{equation}\label{ShapeOfCharacteristics}
\ddot\gamma(s)=\gamma(s).
\end{equation}

\section{Numerical tests}\label{section:numerical_tests}
In this section we perform several numerical tests for the presented models considering different values of parameters, replicating different scenarios: stabilization around an optimal value, cycles (occurring through the evolutionary rescue phenomena) and the extinction.We then compare the numerical results obtained for the stochastic and deterministic approaches, using in particular an asymptotic-preserving scheme which allows us to observe the population dynamics on the passage from the integro-differential equation \eqref{equation:PDE_eps} to a limit \eqref{equation:hamilton_jacobi}.
Throughout this section we define the birth, death rates and the mutation kernel to those given in \eqref{eq:birth_rate}-\eqref{eq:mutation-kernel} respectively, with the parameters fixed throughout all the experiments to $b \equiv 1$, $d_r\equiv 1$, $C\equiv 0.5$ respectively (unless otherwise stated). 

\subsection{Stochastic model}\label{subsubsection:stoch_simulation}
\subsubsection{The scheme}

Our aim is to simulate the population dynamics over a fixed interval $[0,T]$. 
We begin by simulating an initial population of size $N^0$. We assume that the population is normally distributed around the mean trait $x^0_{mean}$ with a standard deviation $\sigma^0$ so that the resulting vector $X^0\in \mathds{R}^{N^0}$. We know that in a time step $\Delta$, an individual can die, give birth, or be a subject to HT. Each event happens according to a certain probability that we compute from the rates. More detailed description of the simulations is provided in Algorithm 1.

Note that in our setting it is possible that $1,2 \text{ or } 3$ events happen within the same time step. Keeping a discretization time step small helps us to keep a biological sense in our simulation: even if the event of horizontal transfer with an "already dead" individual is possible in our setting (if  $T_d\leq T_{HT}\leq \Delta $), this event is extremely rare. 

\begin{algorithm}[h]\label{alg:algorithm_stochastic}
\caption{Population dynamics on time interval $[0,T]$}
\SetAlgoLined
Random initialization of a population $X^0 :=  \mathcal{N}(x_{mean}^0, \sigma^0)\times N^0 $ \;
\While {$i\Delta \leq T$}{
$X^i = X^{i-1}$, $N^{i-1}=size(X^{i-1})$\;
\For { $\forall x\in X^i$}{
$R_b := b(x),\ R_d := d(x){+CN^{i-1}},\  {R_{HT} := \sum_{y\in X^i}h_K(x-y,N^{i-1})}$\;
$T_b := \lambda(R_b),\ T_d := \lambda(R_d),\ T_{HT} := \lambda(R_{HT})$, where $\lambda$ denotes an exponential random law\;
\If {$T_b \leq \Delta $}{
pick up a new trait $z$ from $\mathcal{N}(x,\sigma)$\;
add a new individual with trait $z$ to $X^i$\;
} 
\If {$T_{HT} \leq \Delta $}{pick a trait $y\in X^{i-1}$ according to the law $\frac{h_K(x-y,N^{i-1})}{\sum_{y\in X^i}h_K(x-y,N^{i-1})}$\; remove individual with trait $x$ and add individual with trait $y$;}
\If {$T_{d} \leq \Delta $}{remove the individual with trait $x$ from $X^i$}
}
\Return $X^i$
}
\end{algorithm}

We simulate the population of initial size $N_0 = 10000$ up to time $T=1000$ with $\Delta = 0.01$, with the parameters being defined at the beginning of the section, and $\alpha$ is a Heaviside function. Even if a Heaviside function is not the most easy to analyze when we pass to the deterministic limit of the system (see Subsections \ref{subsection:pde_model} and \ref{subsection:hamilton-jacobi}), we use it for the stochastic simulation, since it is the most straightforward model for HT in biological context, and is much faster to compute than a smooth function.  We fix all constants but $\tau_0$, which regulates the Horizontal Transfer, and study how it affects the dynamics.
Then we plot the density of the population at each moment of time (left side of each Figure): brighter colors on plot mean that there is a big amount of individuals with very similar traits. On the right top and right bottom we plot the normalized population size (ratio between the actual size and the carrying capacity of the system), and the mean trait.  
 
\begin{figure}[ht]
  \centering
  \begin{subfigure}[b]{0.49\linewidth}
    \centering\includegraphics[width=\textwidth]{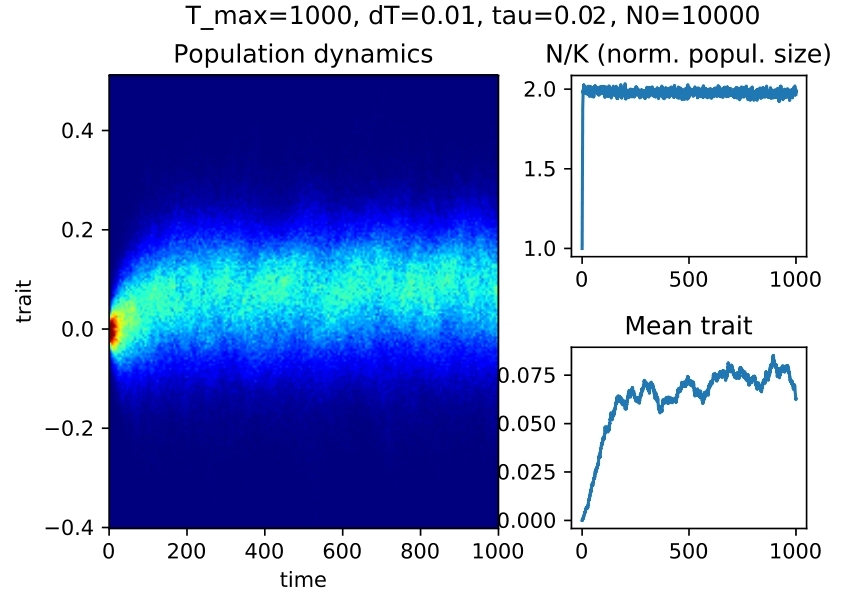}
    \caption{Stabilization: $\tau_0 = 0.02$ }\label{fig:stabilization}
  \end{subfigure}%
  \\
  \begin{subfigure}[b]{0.49\linewidth}
    \centering\includegraphics[width=\textwidth]{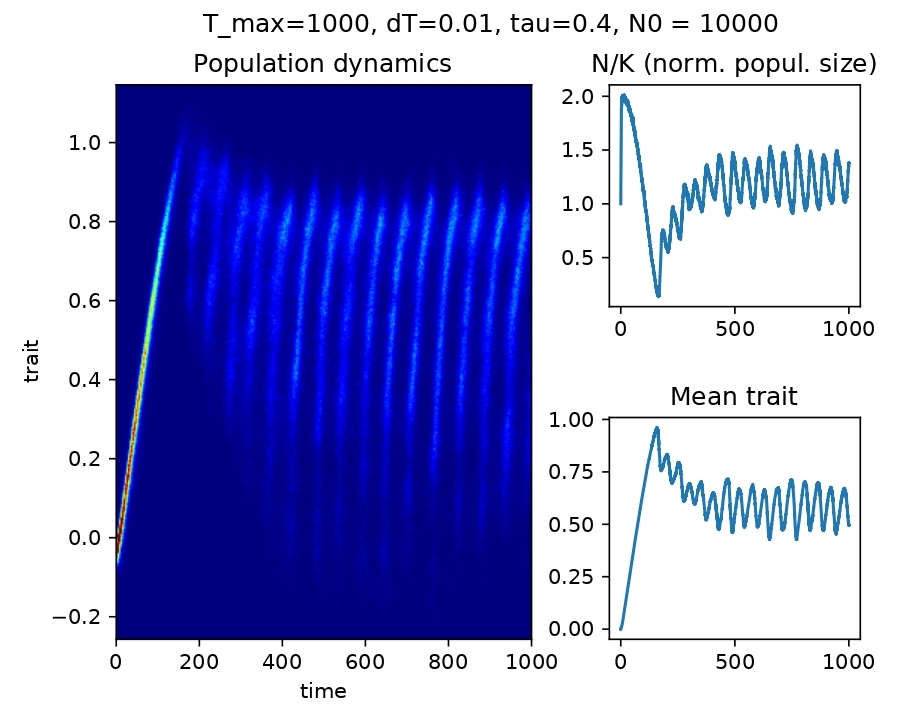}
    \caption{Cycles: $\tau_0 = 0.4$\label{fig:cycles}}
  \end{subfigure}
    \begin{subfigure}[b]{0.49\linewidth}
    \centering\includegraphics[width=\textwidth]{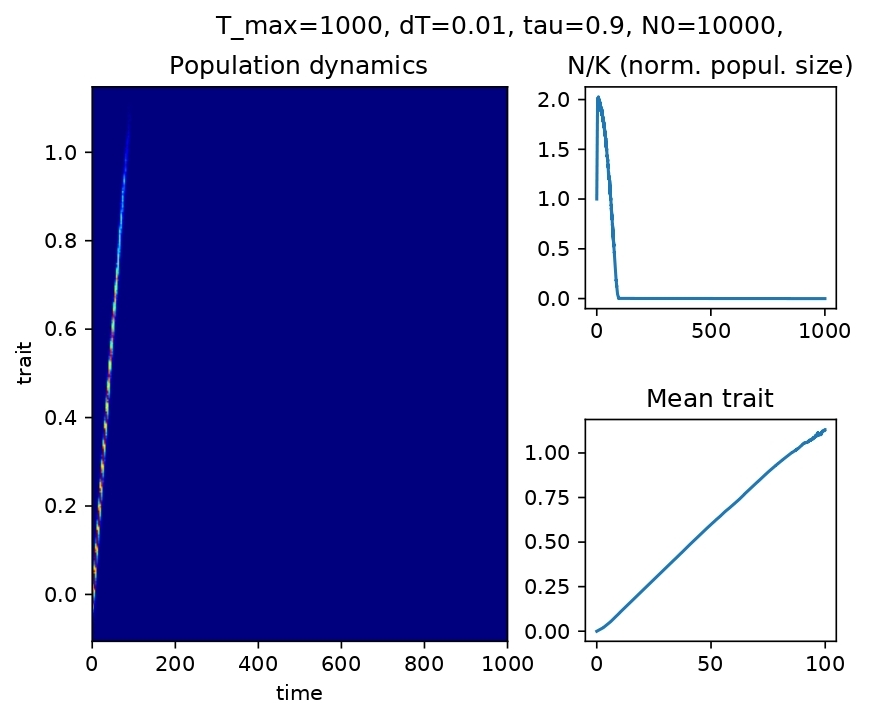}
    \caption{Extinction: $\tau_0 = 0.9$ \label{fig:extinction}}
  \end{subfigure}
\caption{Behavior of the population dynamics as the mutation rate $\tau_0$ is changing, ($b_r = d_r = 1$, $\sigma=10^{-2}$, $K=10^4$, $\sigma^0=10^{-2}$, $x_{mean}^0=0$, $N^0=10^4$).} \label{fig:replication_types}
\end{figure}

Depending on the parameters we may observe three types of behavior (see Figure \ref{fig:replication_types}). First possibility, for small values of $\tau_0$, is the stabilization (Figure \ref{fig:stabilization}).
In this case the population rapidly reaches the equilibrium and concentrates around the optimal trait, which is close to $0.1$
(with stochastic fluctuations). Note that in this case, the mean trait is shifted in comparison to the optimal trait without HT (which is $x=0$).

Second option, for intermediate values of $\tau_0$, is the cycling behavior (Figure \ref{fig:cycles}). Since the transfer rate is sufficiently large, the population is driven towards a deleterious trait, which is eventually less fit than the trait $x=0$. 
If the drift is not too strong, the very few individuals which were not affected by HT and remained fit (with $x$ close to $0$) manage to regrow and eventually repopulate the environment, which launches the cycle again. 

The last possibility, for large values of the horizontal transfer rate $\tau_0$, is the extinction of the population (Figure~\ref{fig:extinction}). It occurs because too many individuals were affected by deleterious traits of their neighbors, so that they die faster than is needed for replicating the population. 

To understand better this phenomenon, we have to give a precise definition of what do we actually refer to, when we say "the critical value" of the transfer rate?  In stochastic setting the answer is not trivial, and that is where the individual-based model reaches its limit. What we observe experimentally is the following when we change the value of HT rate starting from zero, the cycles in the population dynamics become more clearly visible, the fluctuations of the mean trait and the population size become more ample, until at some point the probability of extinction overweights the probability of survival and, finally, at the value of $\tau_0$, which we call "critical" we obtain an almost sure extinction.

\begin{figure}[ht]
  \centering
    \begin{subfigure}[b]{0.49\linewidth}
    \centering\includegraphics[width=\textwidth]{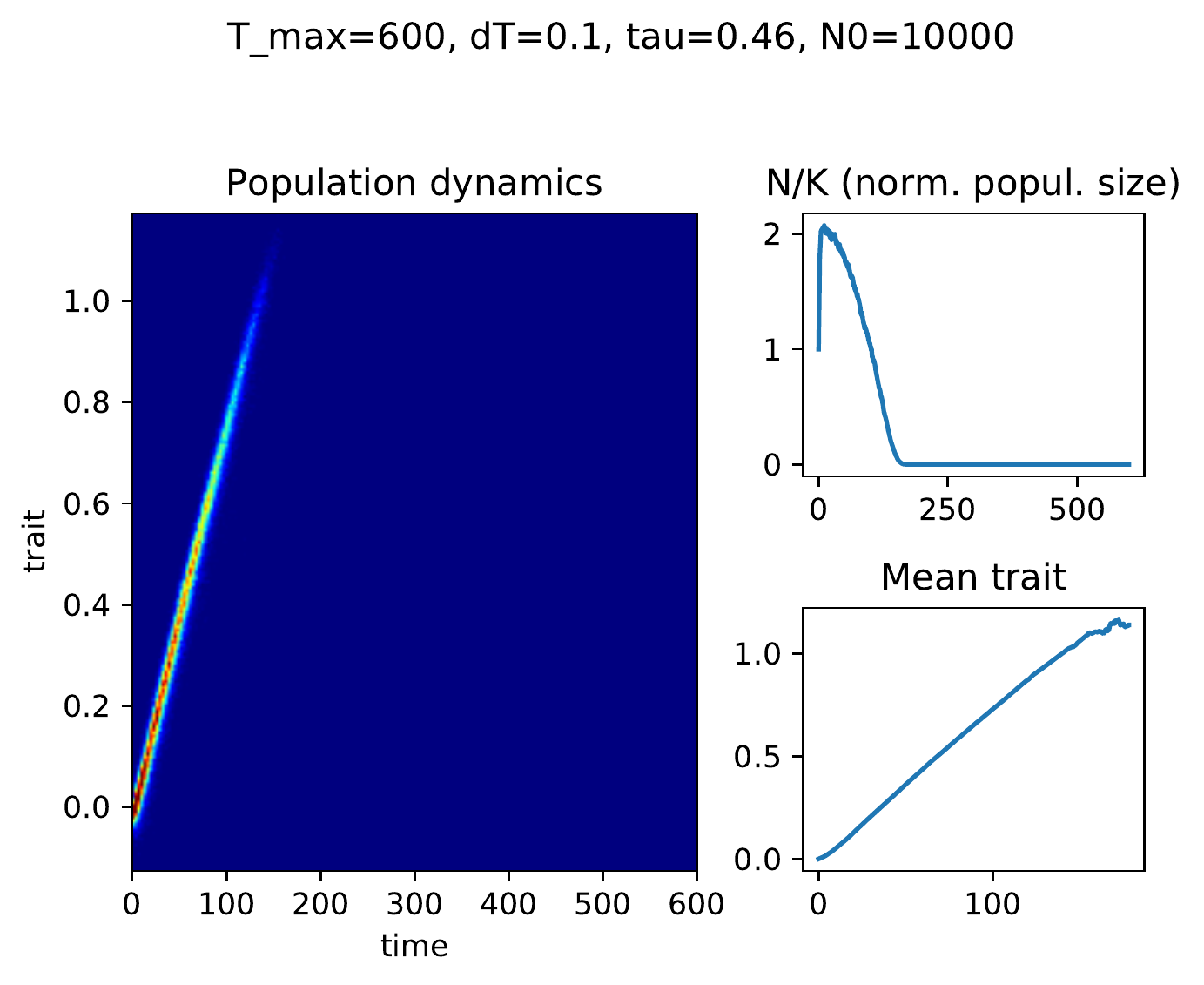}
    \caption{Extinction\label{fig:Behavior_Extinction}}
  \end{subfigure}
\begin{subfigure}[b]{0.49\linewidth}
    \centering\includegraphics[width=\textwidth]{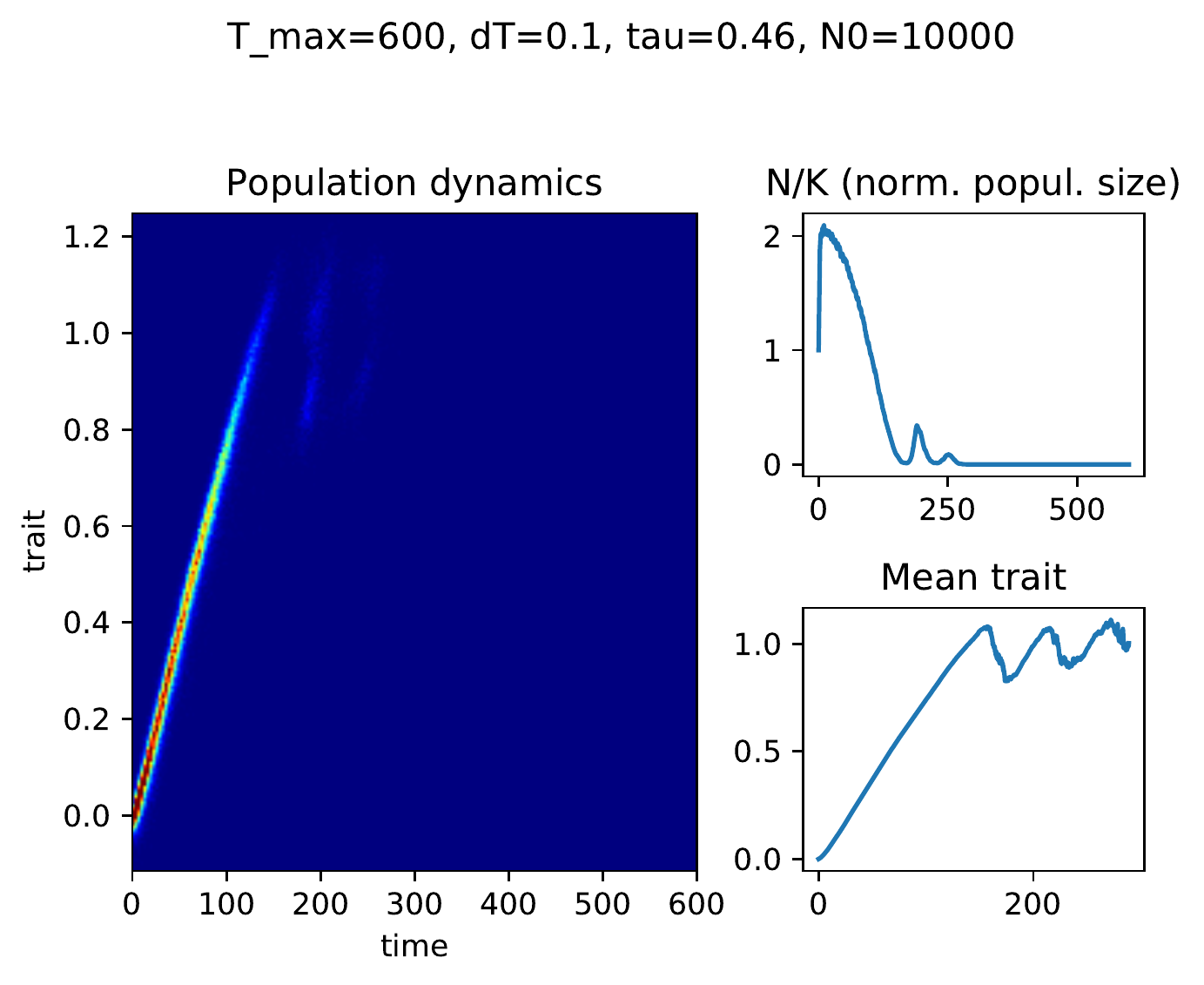}
    \caption{Cycle and extinction}\label{fig:Behavior_Cycle_Extinction}
  \end{subfigure}
  \\
  \begin{subfigure}[b]{0.49\linewidth}
    \centering\includegraphics[width=\textwidth]{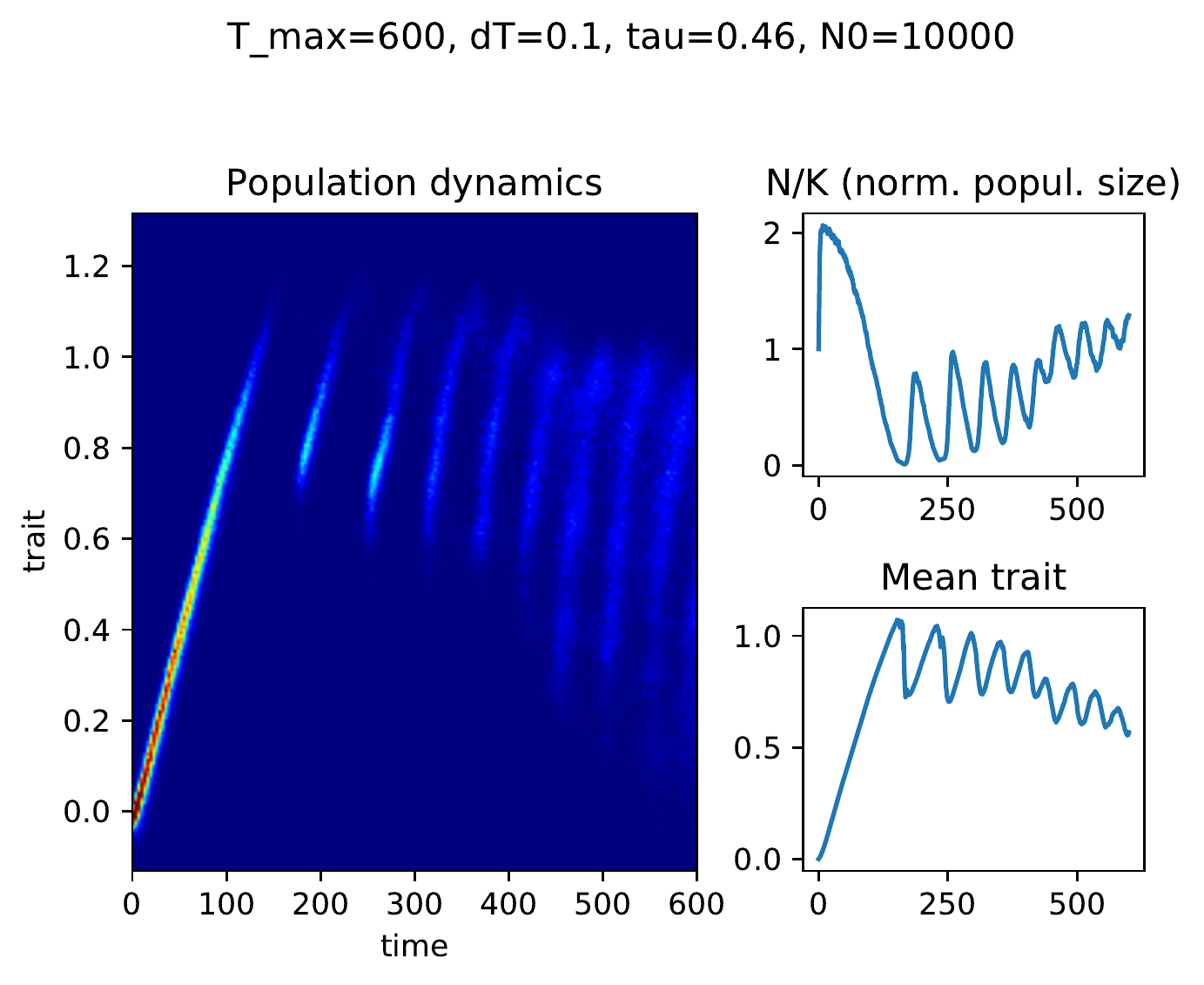}
    \caption{Cycles }\label{fig:Behavior_Cycle}
  \end{subfigure}
  
    \caption{Different behaviors for $\tau_0=0.46$ (and the other parameters as in Figure~\ref{fig:replication_types}).}\label{fig:TwoDifferentBehaviors}
\end{figure}

But since we are working with a point process, giving a strict definition of a "critical value for an extinction" in terms of probability measures seems to be out of reach. Even in the experimental setting this notion is ambiguous: when the value of $\tau_0$ is getting closer to a "critical" (numerically we observe an almost sure extinction at $\tau_0 = 0.49$), in different repetitions of the same experiment we may observe 
different types of behavior: either cycles, or extinction, which occurs after several cycles. It is illustrated on Figure \ref{fig:TwoDifferentBehaviors}, where the computations, launched with exactly the same set of parameters, give very different results. Furthermore, it is not always clear how to differentiate between the stabilization and cycles, especially when the variance of the mutation kernel is large. To the best of our knowledge, there is no straightforward way to analytically measure the probability of each outcome under given initial conditions, which makes the model difficult to analyse.

This constraint of an individual-based model naturally leads us to studying a limiting system described in Subsection \ref{subsection:pde_model}. 

\subsubsection{Lineages}
With the help of the stochastic model we can keep track of the lineage of an individual $i$ which lives at a final observed time $T$. More precisely, we are interested in a history of a phenotype which leads to a long-term survival of an individual. 

\begin{figure}[ht]
\centering
     \begin{subfigure}[b]{0.45\linewidth}
  \centering\includegraphics[width=\textwidth]{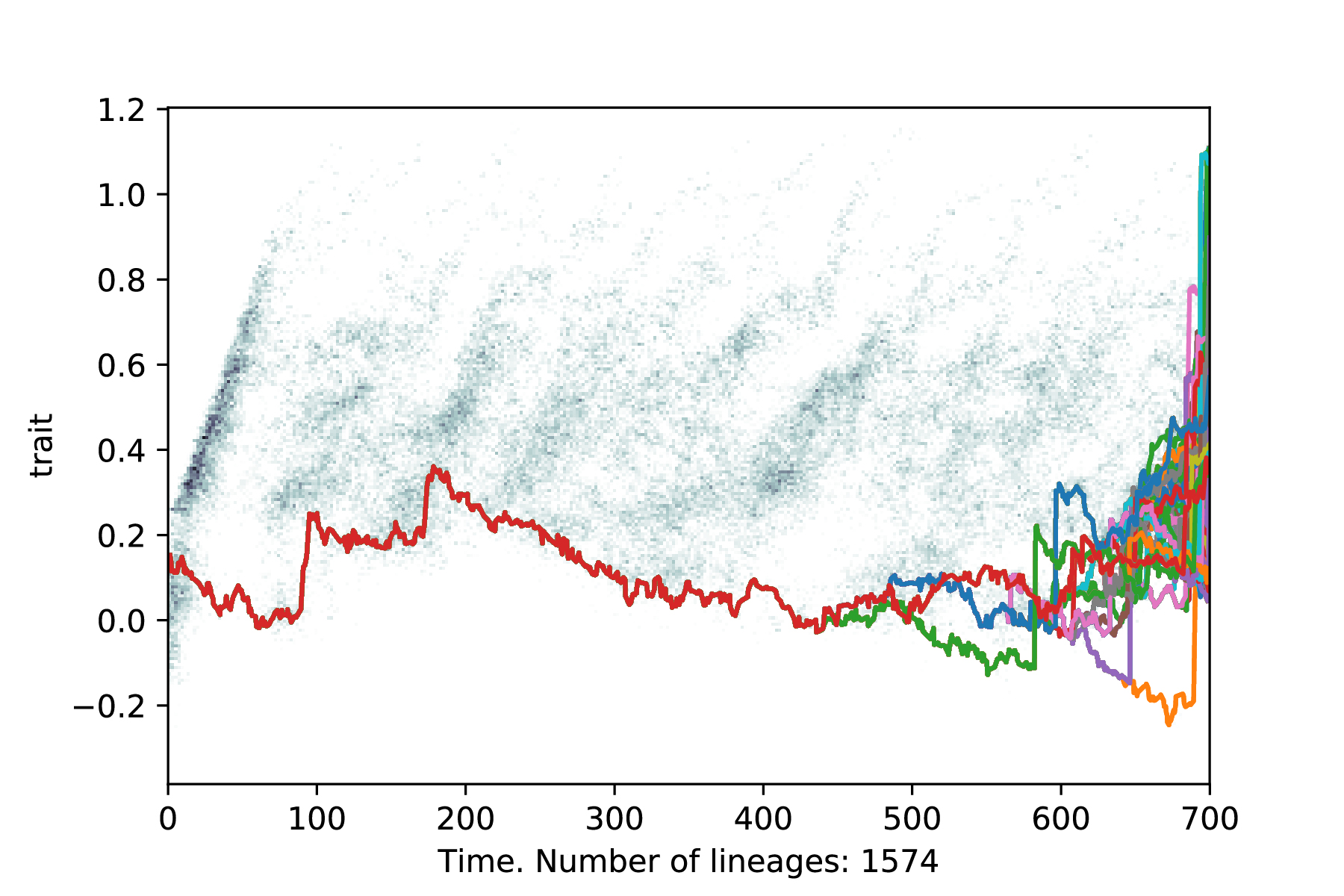}
 \end{subfigure}
      \begin{subfigure}[b]{0.45\linewidth}
  \centering\includegraphics[width=\textwidth]{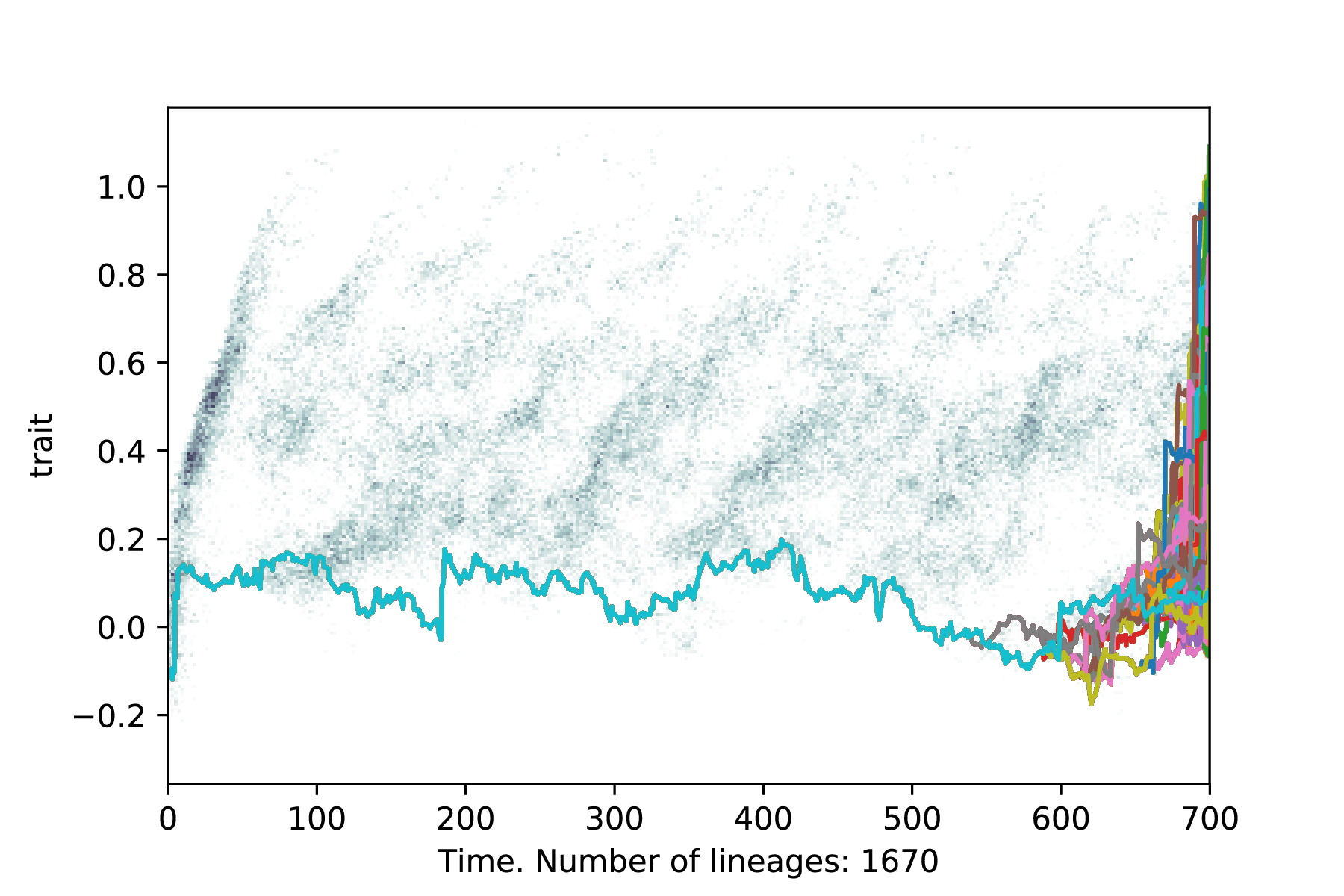}
 \end{subfigure}
      \begin{subfigure}[b]{0.45\linewidth}
  \centering\includegraphics[width=\textwidth]{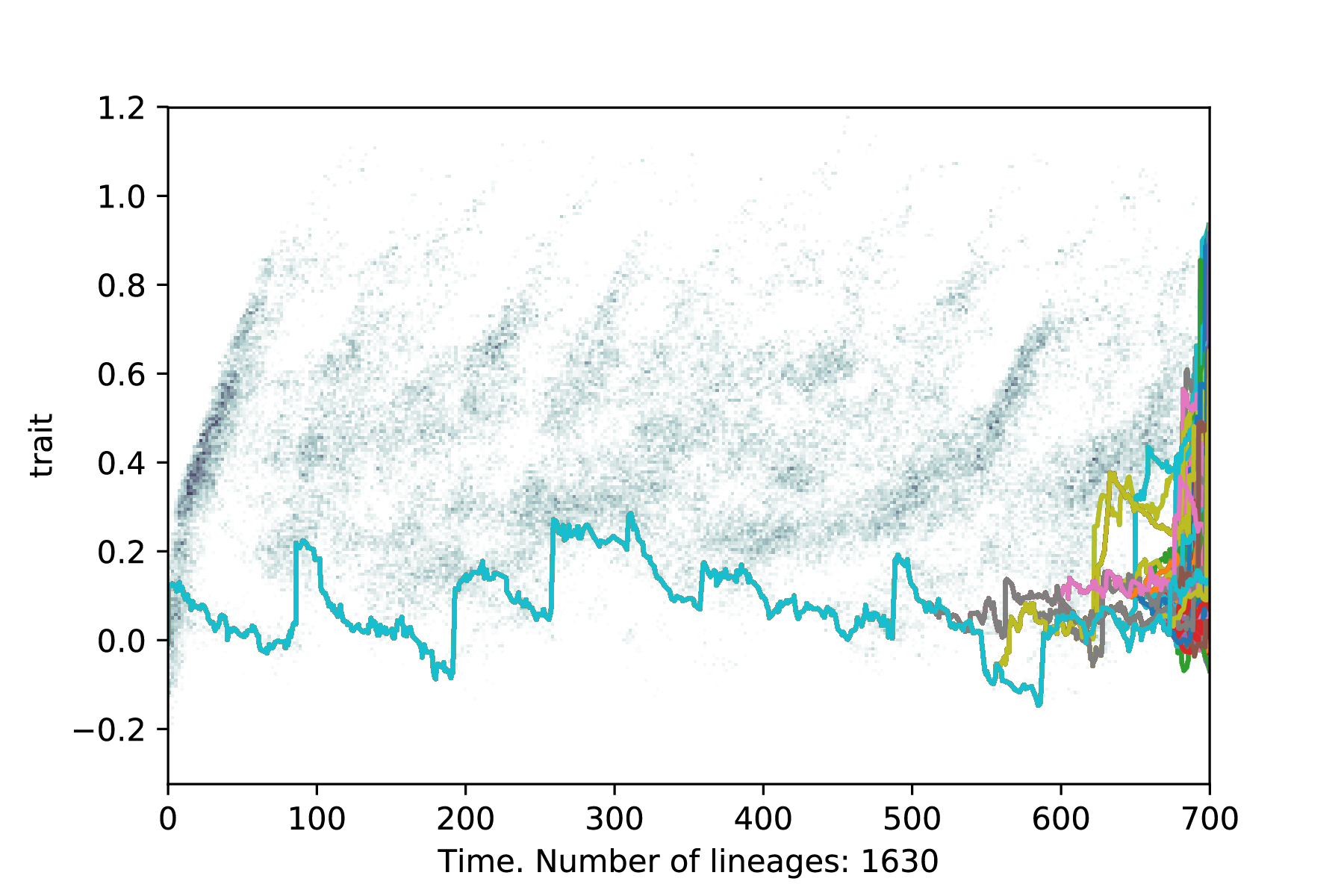}
 \end{subfigure}
      \begin{subfigure}[b]{0.45\linewidth}
  \centering\includegraphics[width=\textwidth]{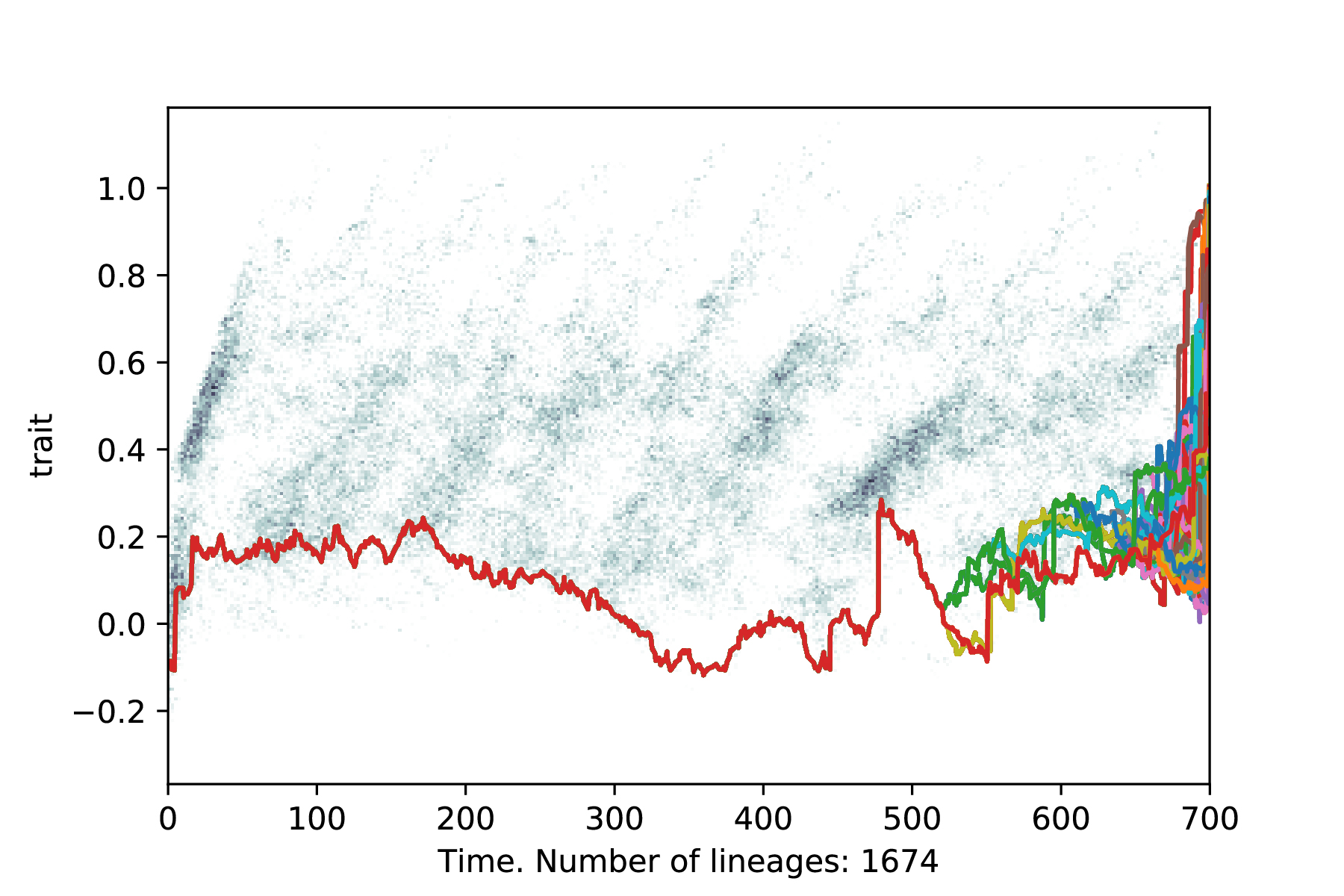}
 \end{subfigure}
 
 \caption{Simulations on the stochastic model with lineages. $\tau_0=0.4$, $T_{max}=700$, $dT=0.1$, $K=N_0=1000$ and other parameters as in Figure~\ref{fig:replication_types}. \label{fig:Lineages}} 
\end{figure}

We illustrate some numerical experiments on Figure~\ref{fig:Lineages}. The four simulations are done with the same parameters. In the background, every point with coordinates $(t,x)$ represents an individual with trait $x$ living at time $t$ (as in Figure 1). The solid lines represent the lineages of the individuals that live at final time. Small fluctuations are the results of birth with mutation, while the large upwards jumps correspond to an occurrence of a HT.

First of all, we can see on the plot that all the lineages are gathered into one line up to $t=400$. It means that all individuals that live at final time $t=700$ emanate from one single ancestor of the initial population. This phenomenon is well known and referred to as \emph{coalescence} in the literature (see for instance \cite{Kingman}, 
or \cite{Arenas&Posada, Arenas429}  for a mathematical description of a classical population genetics theory).

Besides, we see that the lineages remain centered around $x=0$ during almost all the observed time. It is explained by the fact that every lineage that goes to a high value of $x$ (corresponding to deleterious phenotype) cannot recover (since the mutations are small), and eventually goes extinct. This illustrates that the population manage to sustain because of the very few individuals that were not affected by HT throughout the history.

\subsection{
Numerical scheme for the PDE model
}\label{subsection:pde_simul}

In this subsection,  a numerical scheme for \eqref{equation:PDE_eps} is presented, and its properties are numerically investigated. 
For the discretization of \eqref{equation:PDE_eps}, we consider a bounded space of traits $[X_{\mathrm{min}}, X_{\mathrm{max}}]$, discretized with $N_x$ points. Denoting $N_x$ the number of discretization points of the interval $[X_{\mathrm{min}}, X_{\mathrm{max}}]$, we define 
\[
\dx= \frac{X_{\mathrm{min}}-X_{\mathrm{max}}}{N_x-1},
\]
and 
\[
x_i=X_{\mathrm{min}}+i\dx,\;\; 0\le i\le N_x-1.
\]
We consider the time interval $[0,T_{\mathrm{max}}]$, discretized with $N_t$ points $t_n=n\dt$, for $0\le n\le N_t-1$, and where $\dt$ is defined as
\[
\dt=\frac{T_{\mathrm{max}} }{N_t-1}.
\]
The approximations of the solution $f$ of \eqref{equation:PDE_eps} at $(t_n,x_i)$, and of its density $\rho$ at $t_n$ are denoted $f^n_i$ and $\rho^n$ respectively. 
We recall that the initial condition $f^0$ is a smooth function of $x$ given in \eqref{eq:init_condition_f} and the initial density $\rho^0$ is computed using a left-point quadrature rule for $f^0$ as follows: 
\[
\rho^0=\dx\sum\limits_{i=0}^{N_x-1} f^0(x_i).
\]
The scheme is written with an explicit Euler scheme, in which the integrals are computed with a left-point quadrature rule. For $n\ge 1$ and $0\le i\le N_x-1$, it reads
\begin{equation}
\label{eq:scheme}
\varepsilon \frac{f^{n+1}_i-f^n_i}{\dt}= 
\left(d(x_i)+C\rho^n\right)f^n_i  
+ \left[ m*(bf) \right]^n_i + f^n_i \dx \sum\limits_{j=0}^{N_x-1} \tau(x_i-x_j) \frac{f^n_j}{\rho^n}.
\end{equation}

In \eqref{eq:scheme}, the convolution product $\left[ m*(bf) \right]^n_i$ is computed with a left-point quadrature rule, as well of the other integrals.
To do so, a grid in the $z$ variable is defined as for the $x$ variable. Let $Z_{\mathrm{min}}$ and $Z_{\mathrm{max}}$, and the number $N_z$ of discretization  points be given. The grid in $z$ is defined as 
\[
\forall 0\le k\le N_z-1, z_k=Z_{\mathrm{min}}+k\dz,
\]
where $\dz=\left(Z_{\mathrm{max}}-Z_{\mathrm{min}}\right)/\left(N_z-1\right)$. When $x_i+\varepsilon z_k \in \left[X_{\mathrm{min}},X_{\mathrm{max}}\right]$,  the value of $f(t_n,x_i+\varepsilon z_k)$ is approximated by linear interpolation of the $(f^n_i)_{0\le i \le N_x-1}$. When $x_i+\varepsilon z_k< X_{\mathrm{min}}$, or $x_i+\varepsilon z_k> X_{\mathrm{max}}$, it is computed with a linear extrapolation of the $(f^n_i)_{0\le i\le N_x-1}$, using the slope at the corresponding end of the $X$ domain. Using the notation $f^n(x_i+\varepsilon z_k)$ for the approximation of $f(t_n,x_i+\varepsilon z_k)$, we then define
\[
\left[ m*(bf) \right]^n_i= \dz \sum\limits_{k=0}^{N_z-1} m(z_k) b(x_i+\varepsilon z_k) f^n(x_i+\varepsilon z_k). 
\]

\subsubsection{Case $\e = 1$: comparison with stochastic model}

First thing that we are interested in is whether under identical parameters and initial conditions we may reproduce the same
behavior as in the stochastic model. 
Thus, we conduct several experiments, fixing parameter $\e$ to $1$ (thus, we do not rescale time, nor mutation rate), leaving all the other parameters fixed to the same values as in the stochastic simulation case.

\begin{figure}[ht]
  \centering
  \begin{subfigure}[b]{0.45\linewidth}
    \centering\includegraphics[width=\textwidth]{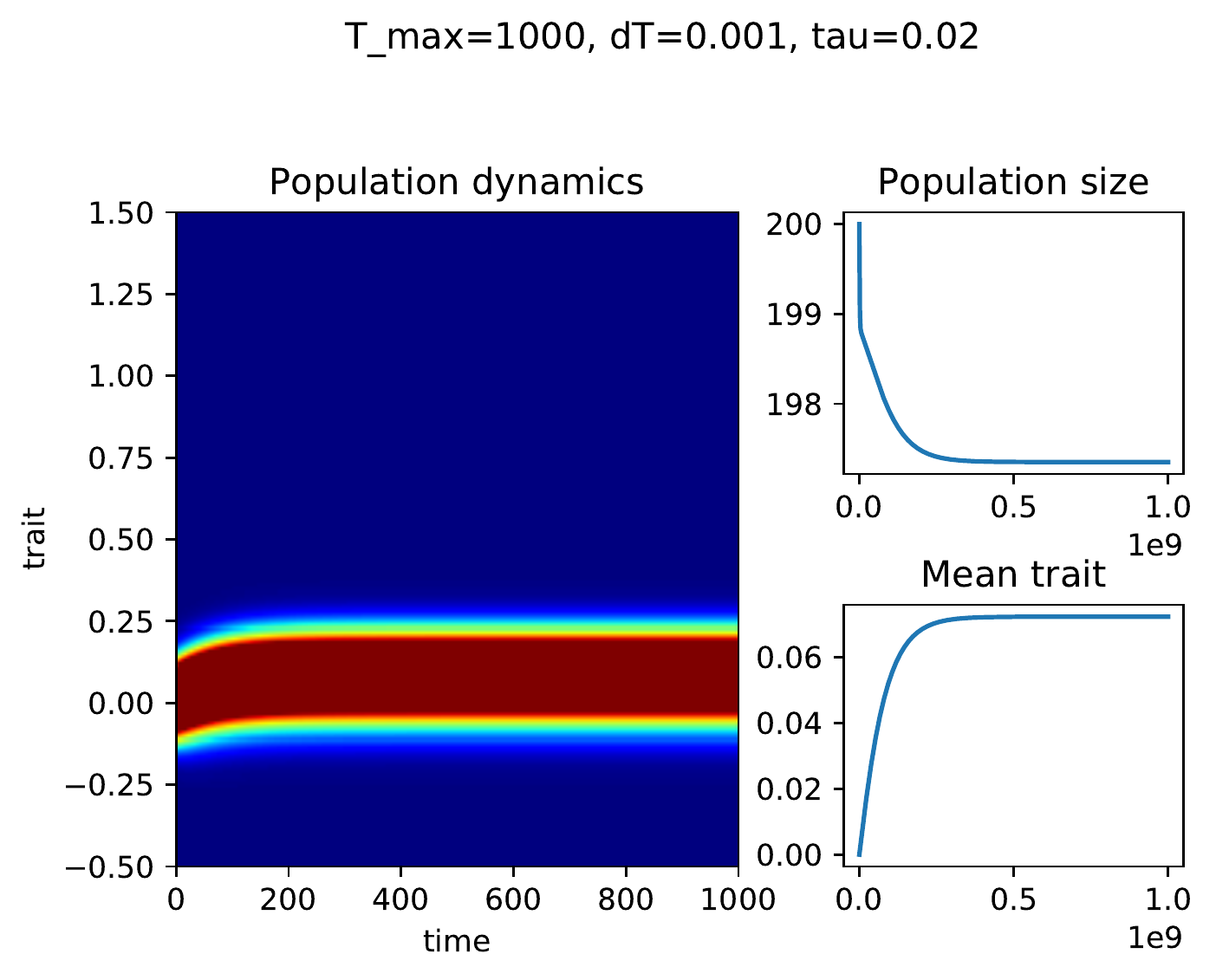}
    \caption{Stabilization: $\tau_0 = 0.02$\label{fig:stabilization_PDE}}
  \end{subfigure}%
  \begin{subfigure}[b]{0.49\linewidth}
    \centering\includegraphics[width=\textwidth]{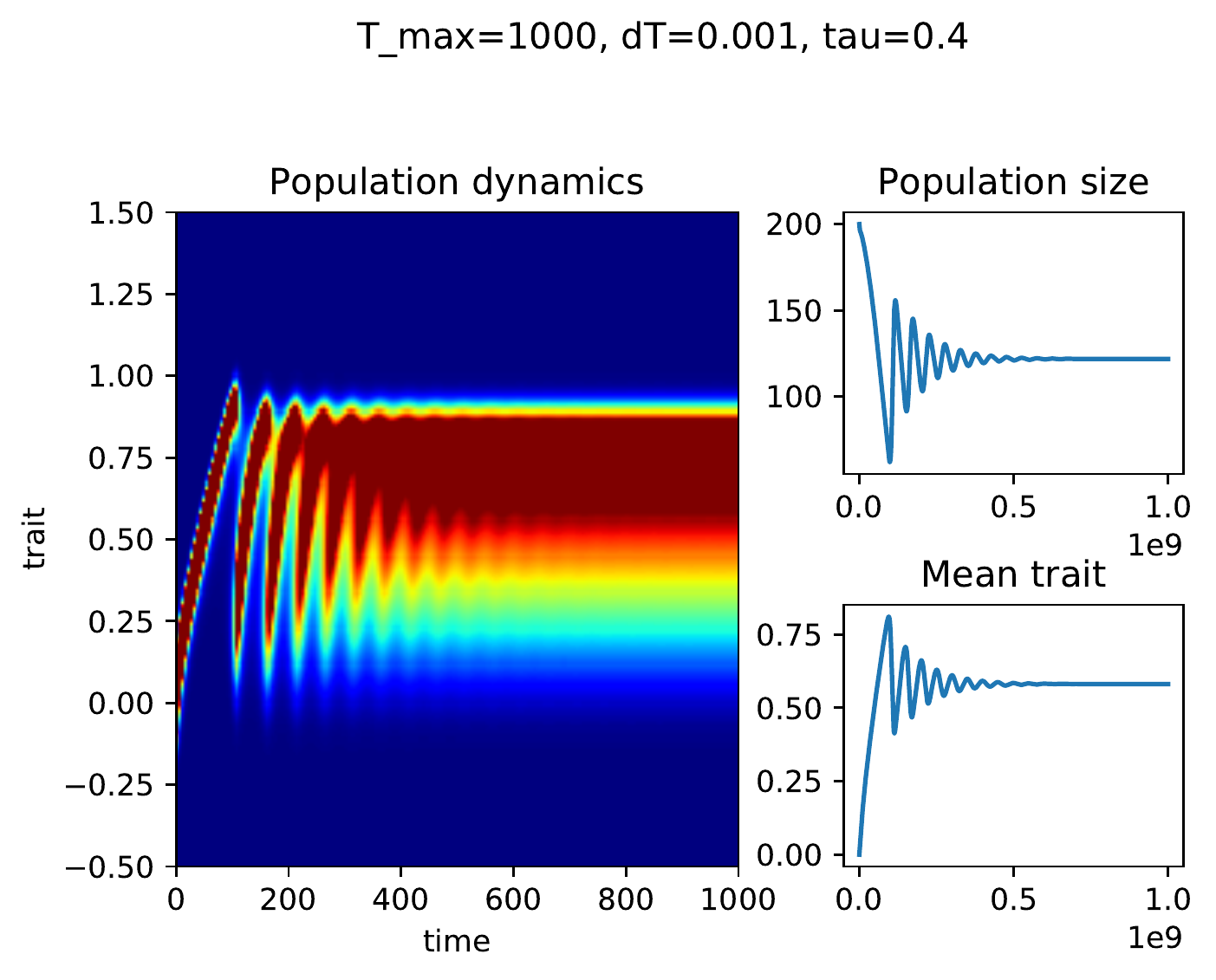}
    \caption{Cycles: $\tau_0 = 0.4$\label{fig:cycles_PDE}}
  \end{subfigure}
 
    \begin{subfigure}[b]{0.49\linewidth}
  \centering\includegraphics[width=\textwidth]{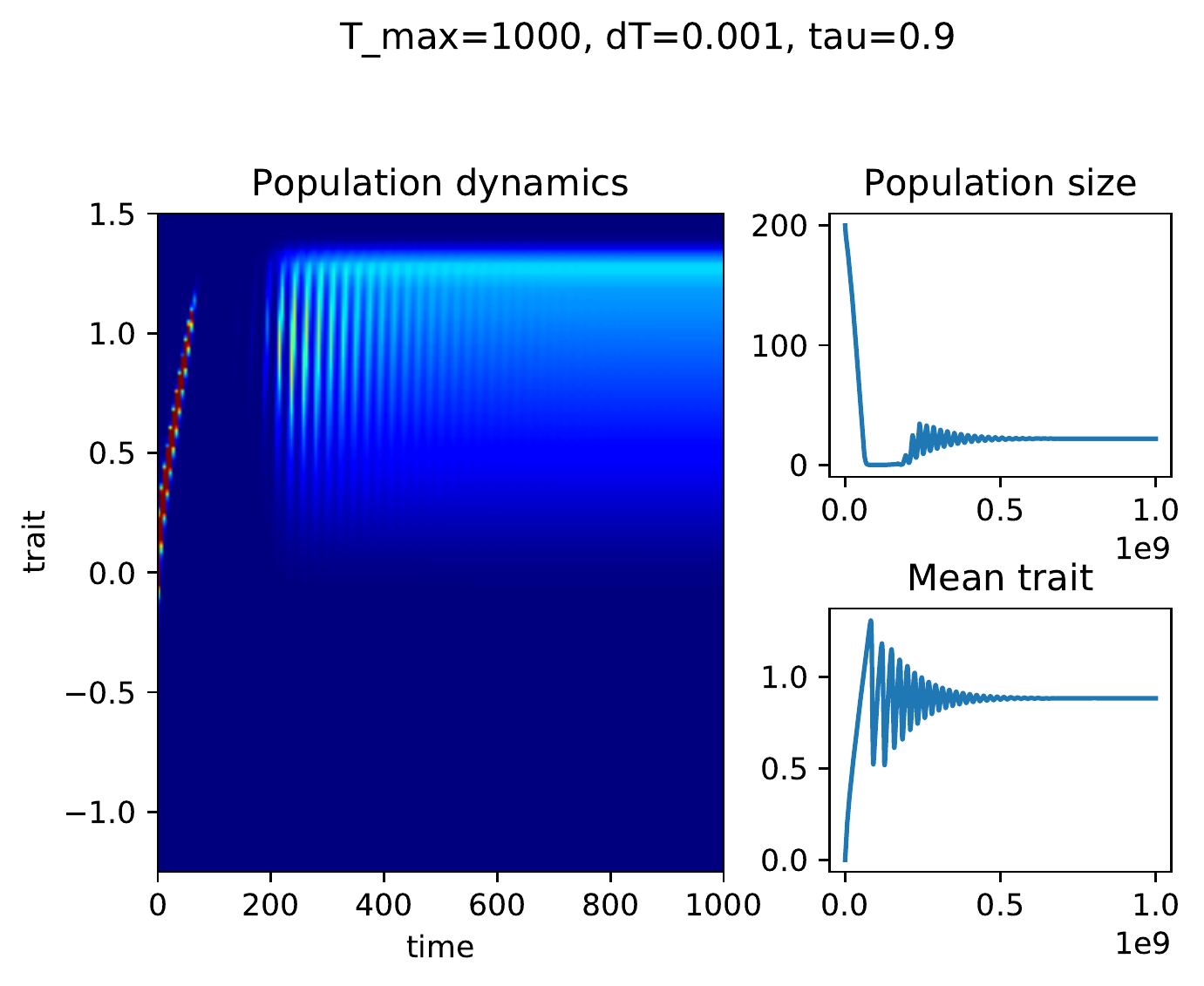}
    \caption{Extinction and cycles: $\tau_0 = 0.9$ \label{fig:extinction_PDE}}
  \end{subfigure}
  \begin{subfigure}[b]{0.49\linewidth}
    \centering\includegraphics[width=\textwidth]{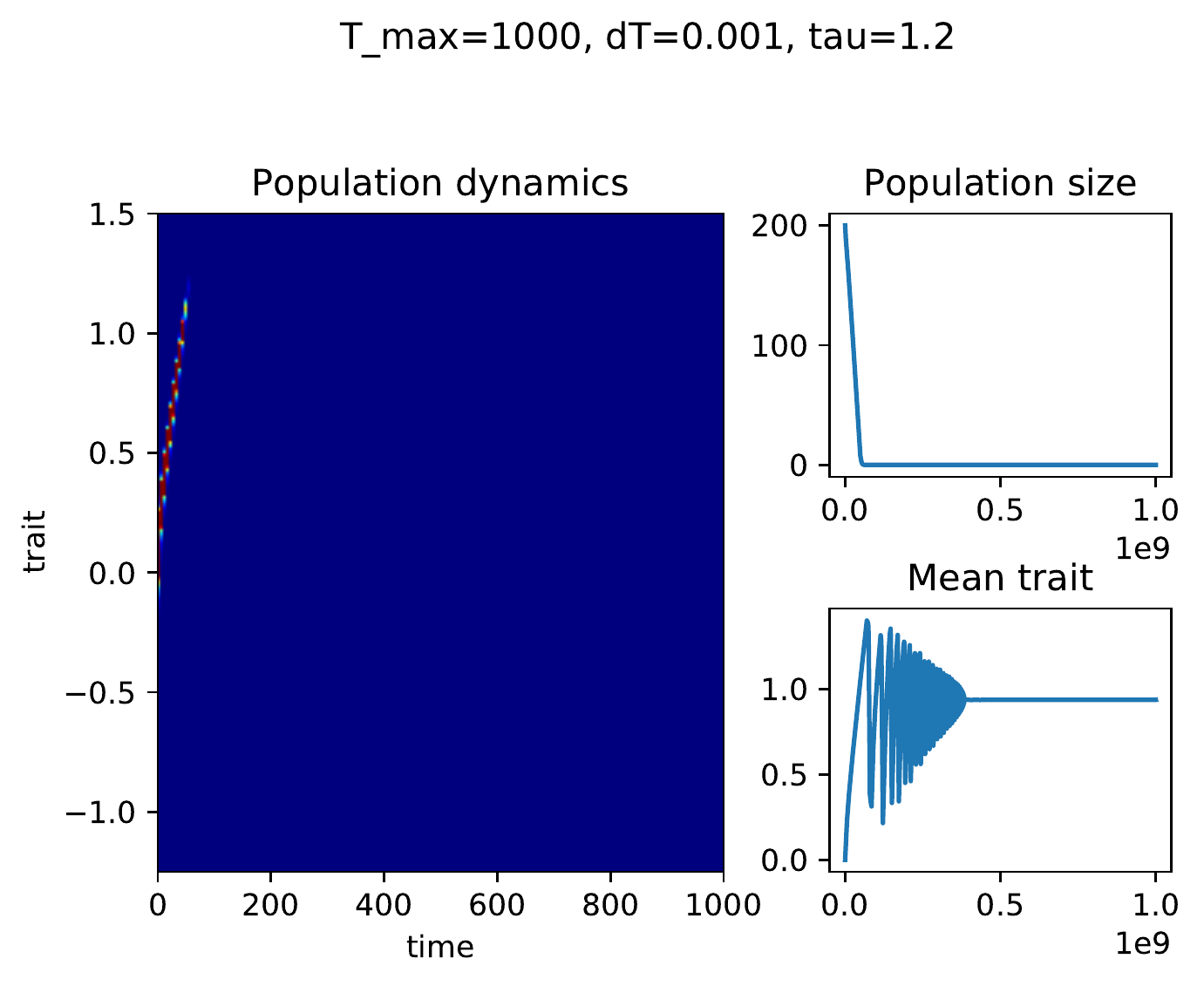}
    \caption{Full Extinction $\tau_0 = 1.2$ \label{fig:Full_extinction_PDE}}
  \end{subfigure}
  
  \caption{Behavior of the population dynamics described by a PDE model as the mutation rate $\tau_0$ is changing, ($b_r = d_r = 1$, $\sigma=0.01$, $\e=1$).} \label{fig:replication_types_PDE}
\end{figure}

As we may see on Figure \ref{fig:replication_types_PDE}, simulations in overall correspond to those of the stochastic model. Indeed, when the HT rate $\tau_0$ is small enough the population rapidly stabilizes around its equilibrium state (see  Figure \ref{fig:stabilization_PDE}), as in the stochastic simulations. Further similarity between two models is that in both cases the optimal trait is shifted a bit above $0$. It is caused by the HT phenomenon.  

For larger values of $\tau_0$, where we would expect to have distinguishable cycles, we observe indeed damped oscillations, see Figure~\ref{fig:cycles_PDE}. We stress out that for the stochastic model it is not the case, see Figure~\ref{fig:cycles}. The way we understand the damping in the oscillations is that the PDE model and the numerical algorithm that we use are not designed to have a precise grasp on the exponential small values of $f$, on which the cycling phenomenon relies. This limitation suggests to perform the change of variable \eqref{eq:u_e_def}, and to write a numerical scheme which converges uniformly when $\e\to 0$. This is what the next subsection is devoted to. 

On Figure~\ref{fig:extinction_PDE}, we observe that as $\tau_0$ becomes larger the population gets extinct, and then, surprisingly enough, "reborns" after a period of extinction. This scenario can only be reproduced on density-based models, since in individual-based model any extinction is definitive. On Figure~\ref{fig:Full_extinction_PDE} we observe a full extinction of the population without regrowth. We will give further insights on those two cases in the next subsection. 

\subsection{The scheme for the Hamilton-Jacobi equation}
\subsubsection{Case $\e\to 0$: description of the numerical scheme}

As the rescaling parameter $\e$ goes to $0$, the model given by \eqref{equation:hamilton_jacobi} gets closer to its limiting state \eqref{HJ_limit}. However, numerical approximation of the \eqref{equation:PDE_eps} for $\e \ll 1$ is not a trivial task. 
 Indeed, for small $\varepsilon$, the solution $f^\ep$ of \eqref{equation:PDE_eps}, is expected to concentrate around the dominant trait. To be able to catch its stiffness numerically, one has to refine the grid in $x$, to ensure enough precision in the computation of $f$. As a consequence, the computational cost of the numerical simulations increases when $\ep \to 0$, and reaching the asymptotic regime with this scheme is not possible. In this part, we present a numerical scheme for \eqref{equation:PDE_eps} which enjoys stability properties in the limit $\varepsilon \to 0$.

To avoid the increase of computational cost when reaching the asymptotics, and to ensure the scheme approaches the limit Hamilton-Jacobi equation for small $\ep$, a scheme for the solution $u^\ep$ of \eqref{equation:hamilton_jacobi} which enjoys the Asymptotic Preserving (AP) property is proposed here. Such schemes have been introduced in \cite{Klar1, Klar2, Jin}, their properties are often summarized by the following diagram: 
\[
\begin{array}{c c c}
 P_\ep &  \xrightarrow[{}]{~~~~\ep\to 0~~~~}  & P_0 \vspace{4pt}   \\
\rotatebox{90}{$\xrightarrow[{}]{~~~h\to 0~~~}$}   
& &\rotatebox{90}{$\xrightarrow[{}]{~~~h\to 0~~~}$} 
\\
 S_\ep^{h} & \xrightarrow[{}]{~~~~\ep\to 0~~~~} & S_0^{h} 
\end{array}
\]
It should be understood as follows: when the parameter $\ep>0$ is fixed, the scheme $S^h_\ep$ is consistent with the $\ep$-dependent problem $P_\ep$. When $\ep$ goes to $0$, the solution of $P_\ep$ converges to the solution of the limit problem $P_0$. The AP scheme $S^h_\ep$ is stable along the transition to the asymptotic regime. It means that, when $\ep$ goes to $0$ with fixed discretization parameters $h$, the scheme becomes a limit scheme $S^h_0$, which is consistent with the limit problem $P_0$.

As an AP scheme is required to enjoy stability properties when $\ep$ is going to $0$, one has to ensure that all the quantities that have to be computed enjoy this property. In the case we are considering, the main concerns are the computation of the integral containing the birth term, the computation of the integral containing the transfer term and the computation of $\rho$. If all of them are correctly defined, the scheme we propose reads 
\begin{equation}
\label{scheme:AP}
\frac{u^{n+1}_i-u^n_i}{\dt}=-(d(x_i)+C\rho^{n+1}) 
+ B^{n}_i + T^{n}_i,
\end{equation}
where $B^{n}_i$ stands for an approximation of 
\begin{equation}
\label{AP:Bni}
\int_\R m(z) b(x_i+\ep z) \mathrm{e}^{(u_\ep(t^n,x_i+\ep z)-u_\ep(t^n,x_i))/\ep} \mathrm{d}z,
\end{equation}
and $T^{n}_i$ is for 
\begin{equation}
\label{AP:Tni}
\int_\R \tau(x_i-y) \frac{f(t^n,y)}{\rho(t^n)} \mathrm{d}y.
\end{equation}
Here, we used the notations and discretization grids defined in the beginning of Section \ref{subsection:pde_simul}, and the dependencies in $\ep$ are omitted to simplify the notations.
In what follows, we present how $T^n_i$, $B^n_i$ and $\rho^{n+1}$ can be computed in a way that ensures they are consistent with their definition for fixed $\ep$, that they can be computed with a constant computational cost with respect to $\ep$, and that their asymptotic behavior when $\ep$ goes to $0$ is meeting the continuous one \eqref{HJ_limit}.
\begin{itemize}
\item \textbf{Computation of $T^n_i$.} The direct approximation of \eqref{AP:Tni} with a quadrature rule is consistent for $\ep\sim 1$. However, since $f$ is expected to concentrate when $\ep\to 0$, it lacks precision in the asymptotic regime. Especially, the convergence of $f/\rho$ to a Dirac is not ensured when the integral is approximated directly. Remarking that 
\[
\frac{\displaystyle f^\ep(t^n,y)}{\displaystyle \rho^\ep(t^n)}=\frac{\displaystyle\mathrm{e}^{u^\ep(t^n,y)/\ep}}{\displaystyle\int_\R \mathrm{e}^{u^\ep(t^n,z)/\ep}\mathrm{d}z} 
= \frac{\displaystyle \mathrm{e}^{(u^\ep(t^n,y)-\max\limits_x u^\ep(t^n,x))/\ep} }{\displaystyle \int_\R \mathrm{e}^{(u^\ep(t^n,z)-\max\limits_x u^\ep(t^n,x))/\ep}  \mathrm{d}z },
\]
\eqref{AP:Tni} is computed with a left-point quadrature rule in the integrals of the previous expression. It reads
\begin{equation}
\label{AP:Tni_formula}
T_i^n=\dx \sum\limits_{j=1}^{N_x-1} \tau(x_i-y_j) \frac{\displaystyle \mathrm{e}^{(u^n_j-\max\limits_l u^n_l)/\ep} }{ \displaystyle \dx \sum\limits_{k=0}^{N_x-1} \mathrm{e}^{(u^n_k-\max\limits_l u^n_l)/\ep} }
=\frac{\displaystyle  
\sum\limits_{j=1}^{N_x-1} \tau(x_i-x_j) \mathrm{e}^{(u^n_j-\max\limits_l u^n_l)/\ep}
}{\displaystyle 
\sum\limits_{k=0}^{N_x-1} \mathrm{e}^{(u^n_k-\max\limits_l u^n_l)/\ep}   
}.
\end{equation}
For fixed $\ep$, \eqref{AP:Tni_formula} is consistent with \eqref{AP:Tni}. Since all the arguments of the exponentials are nonpositive, the limit of \eqref{AP:Tni_formula} for small $\ep$ can be read on that expression. Denoting $j_0$ the index such that 
\[
u^n_{j_0}=\max\limits_l u^n_l,
\]
and supposing that there exists a unique such $j_0$, the limit of \eqref{AP:Tni_formula} for small $\ep$ is
\[
\tau(x_i-x_{j_0}).
\]
This is consistent with the last term in the limit Hamilton-Jacobi equation \eqref{HJ_limit}.
\item \textbf{Computation of $B^n_i$.} Once again, the numerical approximation of \eqref{AP:Bni} is done with a 
quadrature
in the integral. Using the notations of  Section \ref{subsection:pde_simul}, a grid in $z$ is defined. The functions $m$ and $b$ are respectively evaluated at $z_k$ and $x_i+\ep z_k$, but the interpolation of $u^n$ at $x_i+\ep z_k$ has to be done with special care to make the scheme enjoy the expected asymptotic behavior. 
Using a left-point quadrature rule, \eqref{AP:Bni} is approximated by 
\begin{align}
&
\dz \sum\limits_{k=0 \atop \ep |z_k| \le dx }^{N_z-1}
m(z_k) b(x_i+\ep z_k) \mathrm{e}^{
z_k\nabla^{\ep,small}_{n,i,k}
} 
\label{AP:Bni_epz}
+ 
\dz \sum\limits_{k=0 \atop \ep |z_k| > dx }^{N_z-1}
m(z_k) b(x_i+\ep z_k) \mathrm{e}^{
z_k\nabla^{\ep,large}_{n,i,k}
}, 
\end{align}
where $\nabla_{n,i,k}^{\ep}$ stands for an approximation of 
\[
\frac{u^\ep(t^n,x_i+\ep z_k)-u^\ep(t^n,x_i)}{\ep z_k}.
\]
In both cases, it is computed with a linear interpolation of the values $u^n_i$. Hence, $\nabla_{n,i,k}^{\ep, large}$ is given by 
\[
\nabla_{n,i,k}^{\ep, large}=\frac{\tilde{u}^n_{i,k}-u^n_i}{\ep z_k},
\]
where $\tilde{u}^n_{i,k}$ is computed as the linear interpolation of $(u^n_i)_{1\le i\le N_x}$ at $x_i+\ep z_k$. If $x_i+\ep z_k < X_{\mathrm{min}}$ or $x_i +\ep z_k > X_{\mathrm{max}}$, the extrapolation is done linearly using the slope at the first or last point of the interval. Since $\ep z_k > \dx$, no stability issue is faced in this computation.  Still using a linear interpolation, when $0< \ep z_k \le\dx$, it is worth noticing that 
\[
\frac{\tilde{u}^n_{i,k}-u^n_i}{\ep z_k}=\frac{u^n_{i+1}-u^n_i}{\dz},
\]
and when $0>\ep z_k \ge -\dx$, 
\[
\frac{\tilde{u}^n_{i,k}-u^n_i}{\ep z_k}=\frac{u^n_i-u^n_{i-1}}{\dx}.
\]
as a consequence, 
we define:
\[
\nabla^{\ep, small}_{n,i,k}=
 \left\{
 \begin{array}{c l}
 \displaystyle \frac{u^n_{i+1}-u^n_i}{\dx}, & \text{\;\;if\;\;}\displaystyle 0<\ep z_k \le \dx   \vspace{4pt} \\
 \displaystyle \frac{u^n_i-u^n_{i-1}}{\dx}, & \text{\;\;if\;\;}\displaystyle -\dx \le \ep z_k < 0 \vspace{4pt} \\
  \displaystyle 0, & \displaystyle \text{\;\;if\;\;}z_k=0.
 \end{array}
 \right.
\]
This definition of $B^n_i$ is consistent with \eqref{AP:Bni}. Moreover,
 when $\ep$ goes to $0$ with fixed numerical parameters, such as $Z_{\mathrm{min}}$ and $Z_{\mathrm{max}}$, the expression $\nabla_{n,i,k}^{\ep, large}$ is not used at all, and
 \begin{multline}\label{AP:Bni0}
 B^n_i \underset{\ep\to 0}{=} B^{n,0}_i= \dz \sum\limits_{k=0 \atop z_k<0}^{N_z-1} m(z_k) b(x_i) \mathrm{e}^{ z_k \frac{u^n_{i}-u^n_{i-1}}{\dx}} + \dz m(0)b(x_i)+\\ \dz \sum\limits_{k=0 \atop z_k >0}^{N_z-1} m(z_k) b(x_i) \mathrm{e}^{z_k \frac{u^n_{i+1}-u^n_i}{\dx}}.
 \end{multline}
\item \textbf{Computation of $\rho^{n+1}$.} 
In \eqref{scheme:AP}, $\rho^{n+1}$ is considered in an implicit way, to make the limit scheme be consistent with the limit equation \eqref{HJ_limit}. Since 
\[
\rho(t)=\int_{\R} \mathrm{e}^{u(t,x)/\ep} \mathrm{d}x,
\]
for $\ep>0$, we define
\[
\rho^{n+1}=\dx\sum\limits_{i=0}^{N_x-1} \mathrm{e}^{u^{n+1}_i/\ep}.
\]
A closed equation on $\rho^{n+1}$ can be deduced from \eqref{scheme:AP}. Indeed, \eqref{scheme:AP} yields
\[
\mathrm{e}^{u^{n+1}_i/\ep}=\mathrm{e}^{-\dt\rho^{n+1}/\ep}\mathrm{e}^{(u^n_i+\dt \left[-d(x_i) +B^n_i+T^n_i \right] )/\ep},
\]
and so 
 \begin{equation}
 \label{AP:eq_rho}
 \rho^{n+1}=\dx\; \mathrm{e}^{-\dt\rho^{n+1}/\ep} \sum\limits_{i=0}^{N_x-1} \mathrm{e}^{
 A^n_i
 /\ep},
 \end{equation}
 where $A^n_i$ denotes $u^n_i+\dt\left( -d(x_i)+B^n_i+T^n_i \right)$ to simplify the notations. Eventually, $\rho^{n+1}$ is the solution of $h(y)=0$, where
 \begin{equation}
 \label{AP:rho_exp}
 h(y)=y\mathrm{e}^{\dt y/\ep }-\dx \mathrm{e}^{A^n_{i_0}/\ep}\sum\limits_{i=0}^{N_x-1}\mathrm{e}^{(A^n_i-A^n_{i_0})/\ep},
 \end{equation}
where $A^n_{i_0}=\max\limits_i A^n_i$ has been taken apart to get an uniform estimate with respect to $\ep$ on the remaining sum. It is also a solution of the equivalent equation
$g(y)=0$, with
 \begin{equation}
 \label{AP:rho_ln}
  g(y)=-\ep \ln(y)-\dt y +\ep\ln(\dx) + A^n_{i_0}+\ep\ln\left( \sum\limits_{i=0}^{N_x-1} \mathrm{e}^{(A^n_i-A^n_{i_0})/\ep}\right).
 \end{equation}
 To find $\rho^{n+1}$, a Newton's method is applied on expression \eqref{AP:rho_exp} or on \eqref{AP:rho_ln}. Both expressions are smooth convex functions of $\rho$, and are equivalent. Hence, the Newton's method converges whatever is used. Nevertheless, it must be chosen with care.  { \eqref{AP:rho_exp} is to be chosen when $\rho^{n+1}$ is close to $0$ (for large values it becomes less accurate), whereas \eqref{AP:rho_ln} is more adapted when $\rho^{n+1}$ is not small, since it is more prone to accumulate numerical errors when $\rho^{n+1}\to 0$.} In the effective implementation of the method, either one formulation or the other is chosen, depending on the values reached during the iterations of the algorithm. Eventually, to ensure the stability of the numerical resolution of \eqref{AP:eq_rho} when $\ep\to 0$, the inverse of the derivatives of $h$ and $g$ are analytically computed and implemented as 
 \[
 \frac{1}{h'(y)}= \frac{\ep}{\ep+\dt}\mathrm{e}^{-\dt y/\ep}, \;\;\;\;
 \frac{1}{g'(y)}=-\frac{y}{\ep+\dt}.
 \]
 Since $y>0$, these two expressions are uniformly bounded with respect to $\ep$ when $\dt$ is fixed. As a consequence, the cost of the numerical resolution of \eqref{AP:eq_rho} does not increase with $\ep$.
\end{itemize}
When $\ep>0$ is fixed, the scheme \eqref{scheme:AP} is consistent with  \eqref{equation:hamilton_jacobi}, since only quadrature formula and interpolation methods have been used to write it. The way all the terms are computed, as well as the numerical resolution of the non-linear equation \eqref{AP:eq_rho}, ensures the stability of the numerical computations in the small $\ep$ regime. Hence, when $\ep\to 0$ with fixed discretization parameters, the scheme \eqref{scheme:AP}
becomes 
\begin{equation}
\label{scheme:limit}
\frac{u^{n+1}_i-u^n_i}{\dt}=-\left( d(x_i)+C\rho^{n+1} \right) + 
B^{n,0}_i
+ \tau(x_i-x_{j_0}),
\end{equation}
where $j_0$ is such that $u^n_{j_0}=\max\limits_{i}{u^n_i}$, and $B^{n,0}_i$ has been defined in \eqref{AP:Bni0}. 

We do not give a strict proof of consistency of this scheme with respect to the limiting Hamilton-Jacobi equation \eqref{HJ_limit}, since it is out of scope of the project. However, we draw the attention to few important points which need to be taken into account while working with the scheme. In particular, the behaviour of the quantity $\rho(t)$ is not well understood in the case of an extinction. The problem is that intuitively $\rho(t)$ must represent the density of the population --- so that when it goes to zero, we expect an extinction. However, in a Hamilton-Jacobi case even when the $\rho(t)$ reaches zero, the population can still regrow after some time. This can be explained by the fact that after two limiting procedures (passing first to the infinite system size, and then to the infinite time horizon), the "size" of the population can not be described straightforwardly. Accurate link between the quantities obtained as a result of stochastic and PDE simulation is also a question which requires further investigation when $\rho(t)\ll 1$. 

\subsubsection{Case $\e\to 0$: the numerical results}
In this subsection we simulate the dynamics of the population by considering a small value of $\e$ and discuss the obtained results in order to compare them with previous simulations. Note that, in order to compare both, the stochastic and the Hamilton-Jacobi behaviours, the first thing to do is to obtain the simulations for the stochastic model also in the case where the HT rate is a smooth function as we do for the Hamilton-Jacobi case. We recall that, in subsection \ref{subsubsection:stoch_simulation} simulations for stochastic model are done with a Heaviside function as HT rate since it is a more natural choice for simulation of a jump process.

\begin{figure}[ht]
  \centering
  \begin{subfigure}[b]{0.49\linewidth}
    \centering\includegraphics[width=\textwidth]{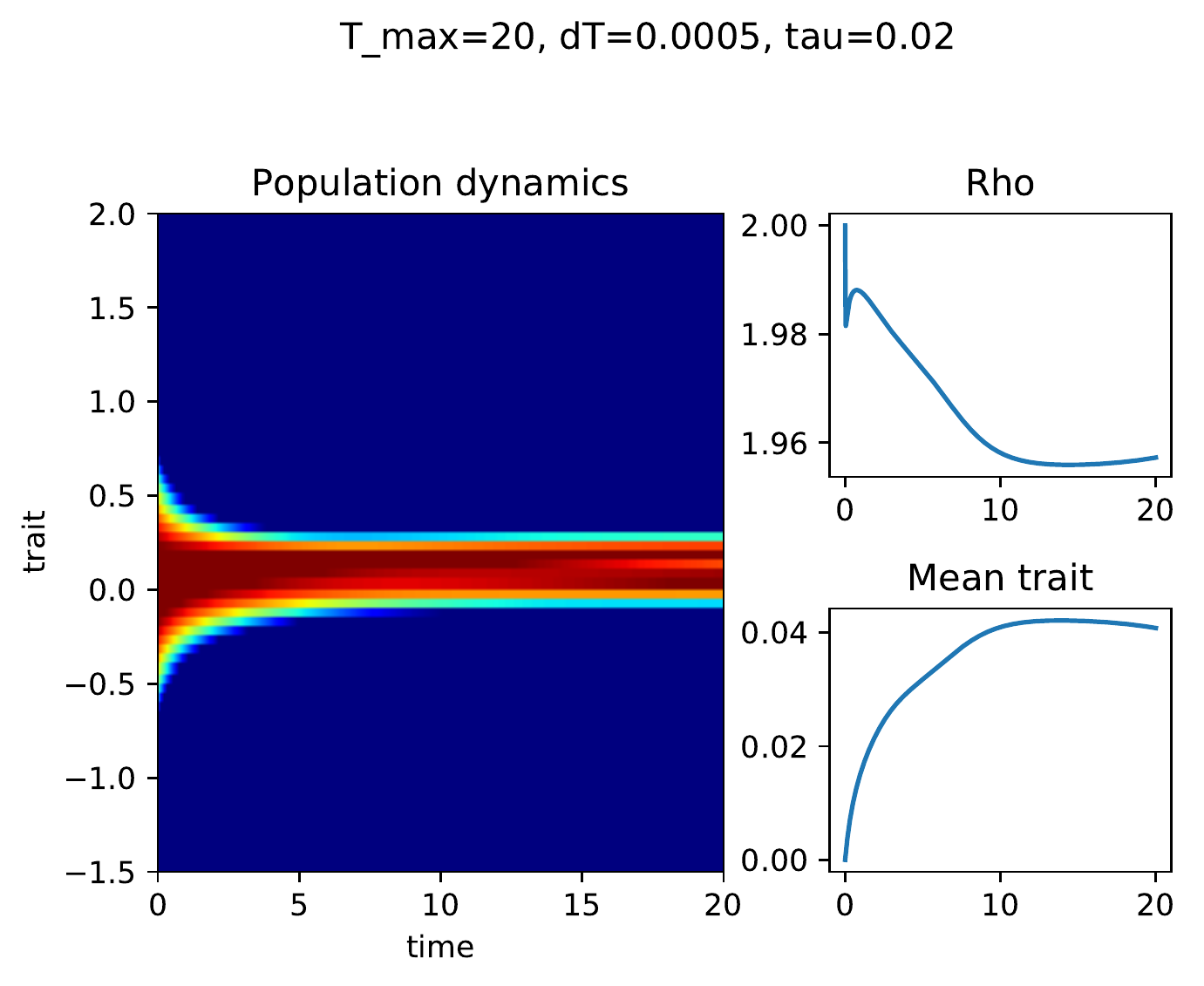}
    \caption{Stabilization: $\tau_0 = 0.02$\label{fig:stabilization_HJ}}
  \end{subfigure}%
  \begin{subfigure}[b]{0.49\linewidth}
    \centering\includegraphics[width=\textwidth]{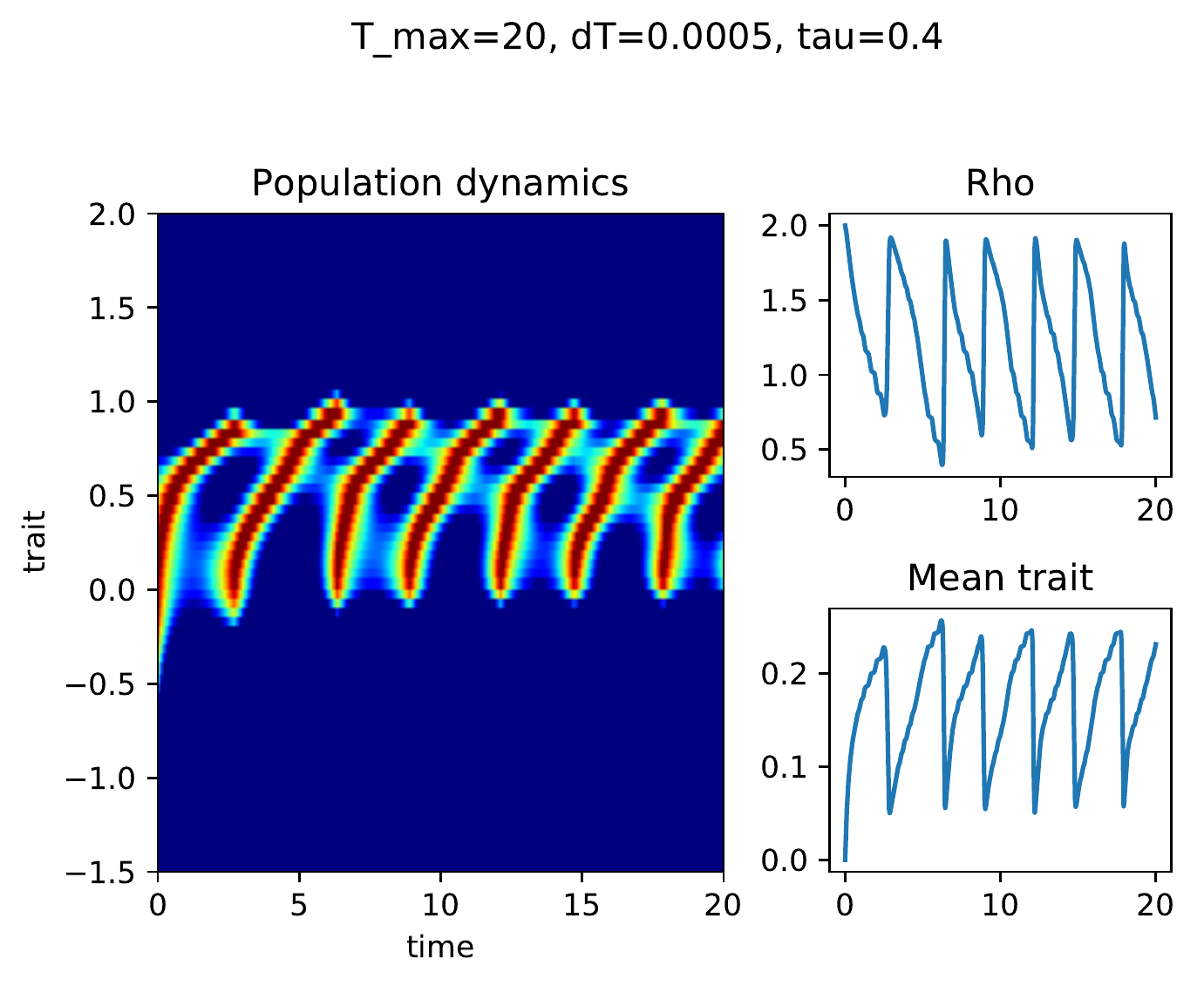}
    \caption{Cycles: $\tau_0 = 0.4$\label{fig:cycles_HJ}}
  \end{subfigure}
    \begin{subfigure}[b]{0.49\linewidth}
    \centering\includegraphics[width=\textwidth]{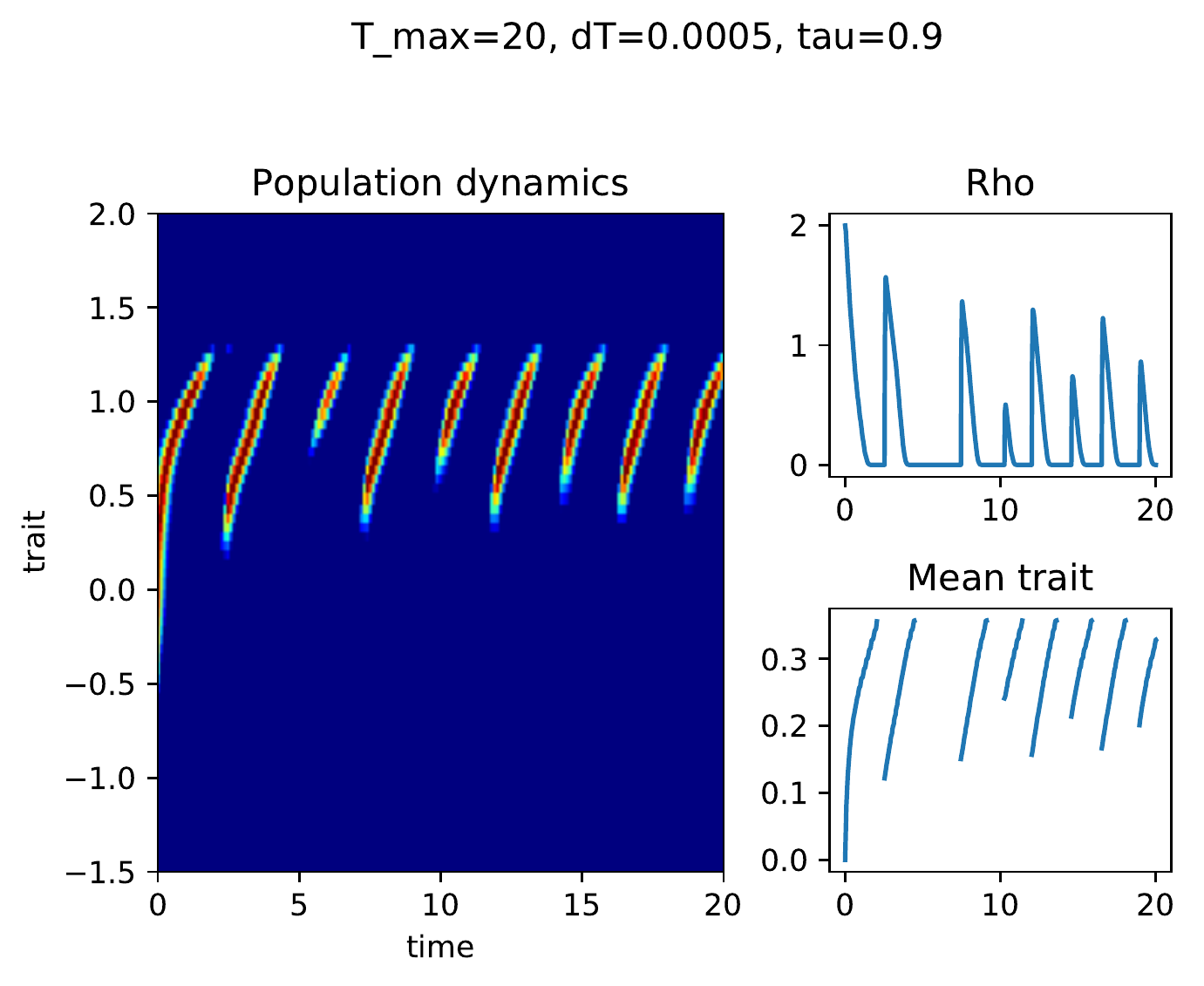}
    \caption{Cyclic extinction: $\tau_0 = 0.9$ \label{fig:extinction_HJ}}
  \end{subfigure}
  \begin{subfigure}[b]{0.49\linewidth}
    \centering\includegraphics[width=\textwidth]{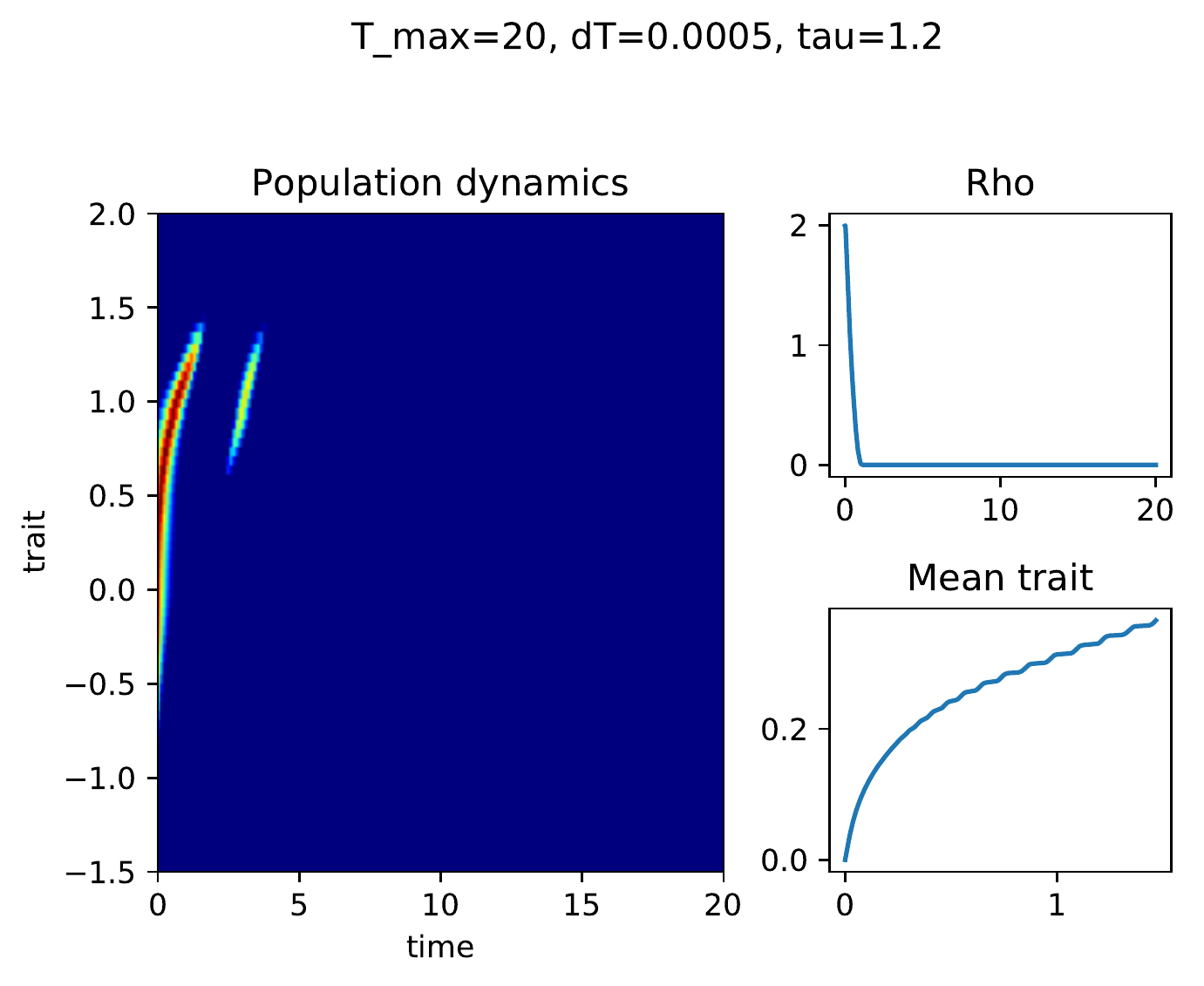}
    \caption{Full extinction: $\tau_0 = 1.2$ \label{fig:full_extinction_HJ}}
  \end{subfigure}
  
  \caption{Behavior of the population dynamics described by a PDE model for $\e = 0.01$ as the mutation rate $\tau$ is changing, ($b_r = d_r = 1$, $\sigma=1$). }\label{fig:replication_types_HJ}
\end{figure}

On Figure \ref{fig:replication_types_HJ} we simulate the population dynamics for $\e = 0.01$. Upon rescaling time (for chosen $\e$ time scale $T=10$ corresponds, in fact, to $\frac{T}{\e}=1000$ in previous simulations) and the variance parameter, we see the same patterns, with few differences. 

On Figure~\ref{fig:stabilization_HJ}, we observe a stabilization of the mean trait, as in Figure~\ref{fig:stabilization}. Similarly, on Figure \ref{fig:cycles_HJ}, we observe cycles, but on the contrary to the PDE model oscillations are not damped. Moreover, it is worth pointing out that the duration of a cycle here corresponds to what we observe in the corresponding stochastic plot (on Figure~\ref{fig:cycles}) multiplied by $\e=0.01$. 
On Figure~\ref{fig:extinction_HJ}, we also observe a cycling behavior, but the population goes periodically extinct (i.e the population reaches exponentially small value, of order $e^{1/\e}$), and then reborn. On the stochastic model, it corresponds to what is illustrated in Figure~\ref{fig:TwoDifferentBehaviors}. It is not surprising that this behavior is difficult to observe on the stochastic model, since very small populations are likely to go extinct.

On Figure~\ref{fig:full_extinction_HJ}, we can see that the population goes completely extinct. The most interesting case to comment is probably the "partial" extinction seen on \ref{fig:extinction_HJ}. Note that despite the fact that $\rho$ remains at $0$ for some time, the population regrows. The point is that, as it was already mentioned above, this numerical parameter has no 1:1 correspondence to the population size parameter $\frac{N_t}{K}$ used in stochastic model. Also note that similar behaviour of stochastic and HJ model are reproduced under a bit different values of parameters. It is caused by the rescaled time and mutation kernel, so that the rigorous link between two models is still to be developed. 

Another interesting thing to comment is that on Figure \ref{fig:cycles_HJ} we may notice that, from the dynamics of the mean trait and the density of the population, it is easy to estimate the periods of the system. Indeed, since the system is deterministic, we just have to compute the distances between local maxima on each curve. For the stochastic system this task is more difficult, especially for a small population, because it includes filtering problem of a noisy signal. To get more accurate results in stochastic model we have to increase the time scale and number of individuals, which is costly from computational point of view. However, if our goal is to study numerically the lineages which lead to the evolutionary rescue of the population, it is still more straightforward to use the individual-based model.

To finish with, let us give some flavor on the computational cost of the simulations for each type. In Table \ref{tab:computational_time} we give a short overview of the elapsed time for the same values of parameters, but for different schemes. As expected, individual-based model is the most expensive to compute. All the computations were performed in \texttt{numpy} library of Python on MacBook Pro (Intel Core i5 processor, 2,7GHz). 

\begin{table}[h]
\centering
\begin{tabular}{r|c|c}
& $\Delta=0.1, \: T=10$ & $\Delta=0.01, \: T=10$ \\
\hline
SM ($N=1000$) & 3.883s & 38.145s \\
SM ($N=10000$) & 15.805s & 153.255s \\
PDE ($\e=1$)  & 0.186s & 1.673s \\
HJ ($\e=10^{-2}$)  & 0.191s & 1.636s \\
HJ ($\e=10^{-6}$)  & 0.195s & 1.656s \\
\end{tabular}
\caption{Elapsed time for simulation of population dynamics for different models (other parameters are fixed to values used throughout all the other simulations, $\tau=0.5$).}\label{tab:computational_time}
\end{table}

\subsection{Comparison of the theoretical analysis of the Hamilton-Jacobi equation and the numerical simulations of the stochastic model}

\subsubsection{Formal computations}

In this section, we propose some formal computations on the stochastic model, based on the analysis of the Hamilton-Jacobi equation performed in the previous section. To fix ideas, we assume $n=1$ and \eqref{eq:alpha}-\eqref{eq:birth_rate}-\eqref{eq:death_rate}, and we fix all constants but $\tau_0$, as in the previous section. However, we choose the function $\alpha$ as a Heaviside function (this is what has been used in the simulations), which is not a smooth function, and thus will lead to minor modifications compared to the previous section.

We make a strong formal assumption: taking $K\gg1$, we assume that the population behaves like a normally distributed random variable all the time, i.e
\begin{equation*}
\nu_t^K(dx)= \rho(t)\frac{1}{\sqrt{2\pi}s(t)}e^{-\frac{\vert x-\bar x(t)\vert^2}{2s(t)^2}}dx,
\end{equation*}
for some standard deviation $s(t)$ and for $\bar x(t)$ defined in \eqref{definition_x_bar}. We expect $s(t)$ to be of the same order as $\sigma$, but giving a general estimate for $s(t)$ in function of $\bar x(t)$ seems intricate. The normalized size of the population $\rho(t):=\frac{N_t^K}{K}$ is approximately given by (see \eqref{formula_rho})
\begin{equation}\label{formula_rho_stoch}
\rho(t)= \frac{1}{C}r(\bar x(t)),
\end{equation}
where $r$ is defined in \eqref{Definition_r}.

We now formally compute the evolutionary singular state $x_\star$. But as $\alpha$ is a Heaviside function (which formally corresponds to the case when $\delta\to0$ in ~\eqref{Definition_x_star}), our derivations must be slightly adapted. In particular,
$\tau(x-\bar x(t))$ in \eqref{HJ_limit} has to be replaced by
\begin{equation*}
\int_{\R}\tau(x-y)\frac{\nu_t^K(dy)}{\rho(t)},
\end{equation*}
and accordingly, recalling that the weak derivative of a Heaviside is a Dirac mass at $0$,
$\tau'(0)$ in~\eqref{Definition_x_star} has to be replaced by $$\int_{\R}\tau'(\bar x(t)-y)\frac{\nu(dy)}{\rho(t)}=\frac{2\tau_0}{\sqrt{2\pi}s(t)}.$$
We find
\begin{equation}\label{x_star_stoch}
x_\star= \frac{\tau_0}{\sqrt{2\pi}s_\star d_r},
\end{equation}
where $s_\star$ is an unknown corresponding to the standard deviation of the population at equilibrium concentrated at $x=x_\star$. Note that it corresponds to~\eqref{Definition_x_star} with $\tilde\delta:= s_\star\sqrt{\pi/2}$.

We now try to estimate $s_\star$. Formally, $s_\star$ should be such that $u_{\star}(x):=\frac{-(x-x_\star)^2}{2s_\star^2}$  is a stationary solution of~\eqref{HJ_limit}. Differentiating twice, and applying at $x=x_\star$ we find 
$$0=b_r\sigma^2 \left(u_\star''(x_\star)\right)^2-2d_r,$$
(with the reasonable assumption $\tau''(0)=0$), which gives
$$s_\star=\sqrt{\sigma\sqrt{\frac{b_r}{2d_r}}}.$$
Numerically, we find $s_\star= 0.12.$ We end up with the following formula:

\begin{equation}\label{x_star_stoch}
x_\star= \frac{\tau_0}{\sqrt{2\pi\sigma }d_r}\sqrt[4]{\frac{2d_r}{b_r}}.
\end{equation}

\subsubsection{Stabilization}

We run a numerical test on the stochastic model corresponding to stabilization, for $\tau_0=0.02$, and the other parameters as in Figure~\eqref{fig:stabilization}. In this case, $x_\star$ correspond to the mean trait of the population for large time. From, \eqref{x_star_stoch} we find $x_\star=0.067$, and from~\eqref{formula_rho_stoch}, we obtain $\rho_\star=1.99$, which corresponds to what we can see on Figure~\eqref{fig:stabilization}.

\subsubsection{Threshold for cycles}
Since equation \eqref{Definition_x_resc} remains unchanged, we obtain the following threshold for cycles (corresponding to~\eqref{definition_tau_cyc}):
\begin{equation*}
\tau_{cyc}= 2\pi d_r\sigma\sqrt{\frac{b_r}{2d_r}}.
\end{equation*}
With our choice of parameters, we obtain $\tau_{cyc}=0.09$. This threshold 
corresponds to the numerical simulations (however, characterizing precisely whether cycles occurs or not on the numerical simulations is not easy when $\tau_0$ is close to the threshold).

\subsubsection{Threshold for extinction}
Using \eqref{Definition_x_ext}, we can also find a threshold for extinction:
$$\tau_{ext}:=\sqrt{2\pi b_r d_r\sigma}\sqrt[4]{\frac{b_r}{2d_r}}.$$ For our choice of parameters, we obtain $\tau_{ext}=0.30$. 

\begin{figure}[ht]
  \centering
  \begin{subfigure}[b]{0.50\linewidth}
    \centering\includegraphics[width=\textwidth]{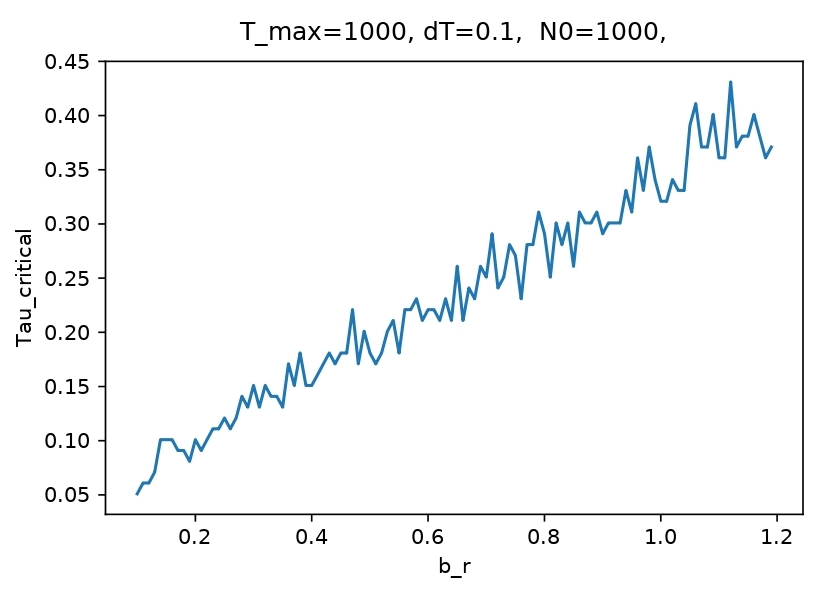}
    \caption{Birth dependency\label{fig:birth_dependency}}
  \end{subfigure}%
  \begin{subfigure}[b]{0.50\linewidth}
    \centering\includegraphics[width=\textwidth]{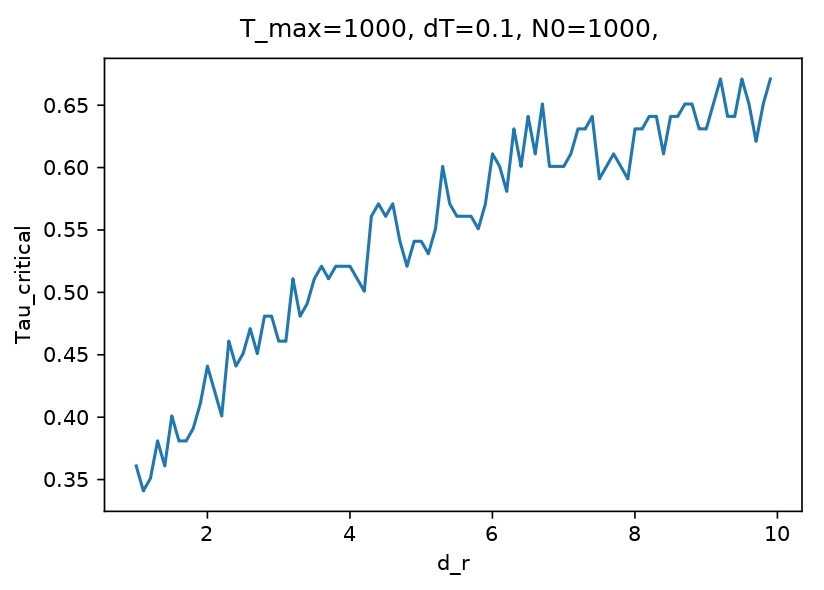}
    \caption{Death dependency\label{fig:death_dependency}}
  \end{subfigure}
 \caption{Dependency on the threshold for extinction $\tau_{ext}$ with respect to the birth rate $b_r$ and death rate $d_r$}\label{fig:critical_tau}
\end{figure}

We now compare this formula with numerical experiments on the individual-based model. They are organized as follows: we fix the birth $b_r$ or the death rate $d_r$, and save the first value of $\tau_0$ under which the extinction occurs. Then, we increase the rate and save the next HT rate under which we have an extinction. Resulting curve for the birth rate is saved on Figure~\ref{fig:birth_dependency} (for death rate: Figure~\ref{fig:death_dependency}). Non-concerned parameters remain fixed as in Subsection \ref{subsubsection:stoch_simulation}. 

The numerical results, in particular, justify at the first glance surprising fact that the extinction threshold depends on the birth and death rate in the same manner. It seems logical to assume that while the higher birth rate contributes to a bigger survival probability even with a relatively big horizontal transfer rate, higher death rate must have an opposite effect. However, in conditions of a very "harsh" environment individuals with non-fit traits die out before they manage to transfer their genetic information to the other individuals. As a consequence, value of the critical $\tau$ increases as the value of the birth (or death) rate constant increases.

\section*{Conclusions}
First achievement of the paper consists in an accurate numerical study conducted on the stochastic model given by a point measure \eqref{eq:point_measure}. To the best of our knowledge, in-depth analysis of the influence of the HT rate on the evolutionary dynamics has not been yet attempted. Along with its accuracy, the stochastic model reveals its limitation: an accurate theoretical description of what actually happens in each observed scenario from a mathematical point of view seems to be out of reach. However, we show that this model can be used for tracing back the lineage of the survived individuals through several generations.

On the next step, in a numerical comparative study between the stochastic (individual based) and the PDE (density) model both models exhibit the same behavior for a given set of parameters, which justifies theoretical results from \cite{Billiard2016,BFRMT_Stochastic}. Minor differences (in particular, the presence of damping oscillations) can be explained by a choice of a numerical scheme. However, further analysis shows that the classical PDE model defined by \eqref{equation:PDE} leads to instabilities if we try to pass to an asymptotic setting under the small mutation assumption. Those instabilities are then resolved by a transformation of an initial model to a Hamilton-Jacobi type equation and using an asymptotic-preserving scheme. Further advantage of this approach is that the resulting equation \eqref{equation:hamilton_jacobi} makes an easier subject of a theoretical analysis. 

\begin{figure}[h!]
\centering
\begin{subfigure}[b]{0.48\linewidth}
\centering\includegraphics[width=\textwidth]{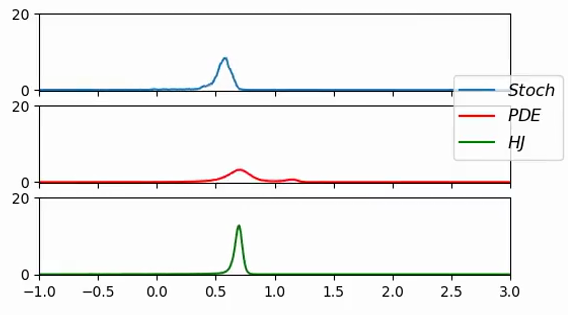}
\caption{$t=167$ \label{fig:SM_1}}
\end{subfigure}%
\begin{subfigure}[b]{0.48\linewidth}
\centering\includegraphics[width=\textwidth]{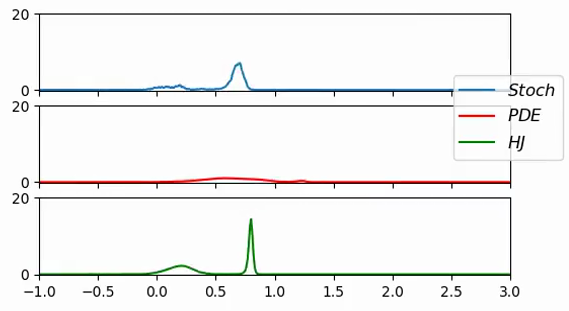}
\caption{$t=198$ \label{fig:SM_2}}
\end{subfigure}%

\begin{subfigure}[b]{0.48\linewidth}
\centering\includegraphics[width=\textwidth]{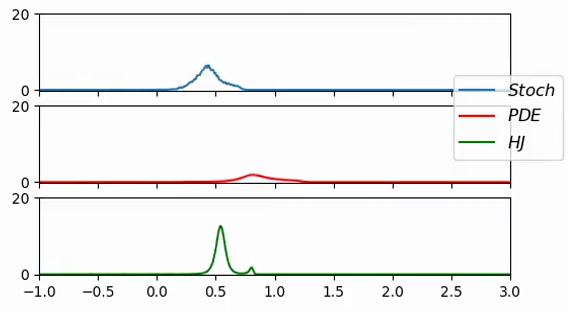}
\caption{$t=216$ \label{fig:SM_3}}
\end{subfigure}
\caption{Comparison of numerical simulations between the different models. $\tau_0=0.4$, $\e=0.1$, $\delta=0.001$ and other parameters as in Figure~\ref{fig:replication_types}. Blue line stands for the stochastic model, red line: for a PDE, green --- for a Hamilton-Jacobi PDE}\label{fig:evolutionary_rescue_SM_and_HJ}
\end{figure}

Finally, in a Hamilton-Jacobi setting we manage to numerically replicate the evolutionary rescue of a small population which we observe in the stochastic model. This phenomena is illustrated for stochastic, PDE and HJ simulation on Figure \ref{fig:evolutionary_rescue_SM_and_HJ}. On Figures \ref{fig:SM_1}-\ref{fig:SM_3}  we trace the moment of the regrowth for different models. Figure \ref{fig:SM_1} show the state of the population at certain moment of time: we see how the individuals are centered around a mean trait. For PDE and HJ model (red and green line respectively) we simply plot the density function, and on the first (blue) plot we approximate a histogram which describes ratio $\frac{N_t}{K}$ sorted by traits in stochastic model.  Stochastic simulations show the evolutionary rescue in more distinct manner: we see how the very small number of non-mutated individuals rescues the whole population from extinction (transition from \ref{fig:SM_2} to \ref{fig:SM_3}). On the contrary, the transition on the PDE model is dumped, and the regrowth is not clearly visible. It is due to, again, numerical instability of the PDE scheme for small values of the density function. Finally, HJ explicitly shows how the cycle occurs: the regrow of the "fit" individuals we see in stochastic plot is reproduced by a change of the maximum point (see again \ref{fig:SM_2} to \ref{fig:SM_3}). 

We highlight again that in order to compare the models on a more applied level, we have to give a formal definition of a quantity represented by $\rho$ in a Hamilton-Jacobian case. In this work we have made few steps toward the theoretical analysis of the limiting equation and an accurate description of each event (evolutionary rescue, extinction, etc) in terms of solutions of a PDE. Even though establishing a rigorous mathematical link between the behavior observed in the individual-based model and the Hamilton-Jacobi equation is out of scope of this project, obtained analytical results already give a flavor of how the analysis of the evolutionary dynamics can be simplified in the future.



\section*{Acknowledgements}
This project has received funding from the European Research Council (ERC) under the European Union’s Horizon 2020 research and innovation program (grant agreement No. 639638). The project has received funding from  the Chair “Modélisation Mathématique et Biodiversité” of Veolia Environnement-Ecole Polytechnique-Museum National d’Histoire Naturelle-Fondation X. 

\bibliographystyle{plain}	
\bibliography{references}

\end{document}